# EAISR International Academic Conferences Proceedings

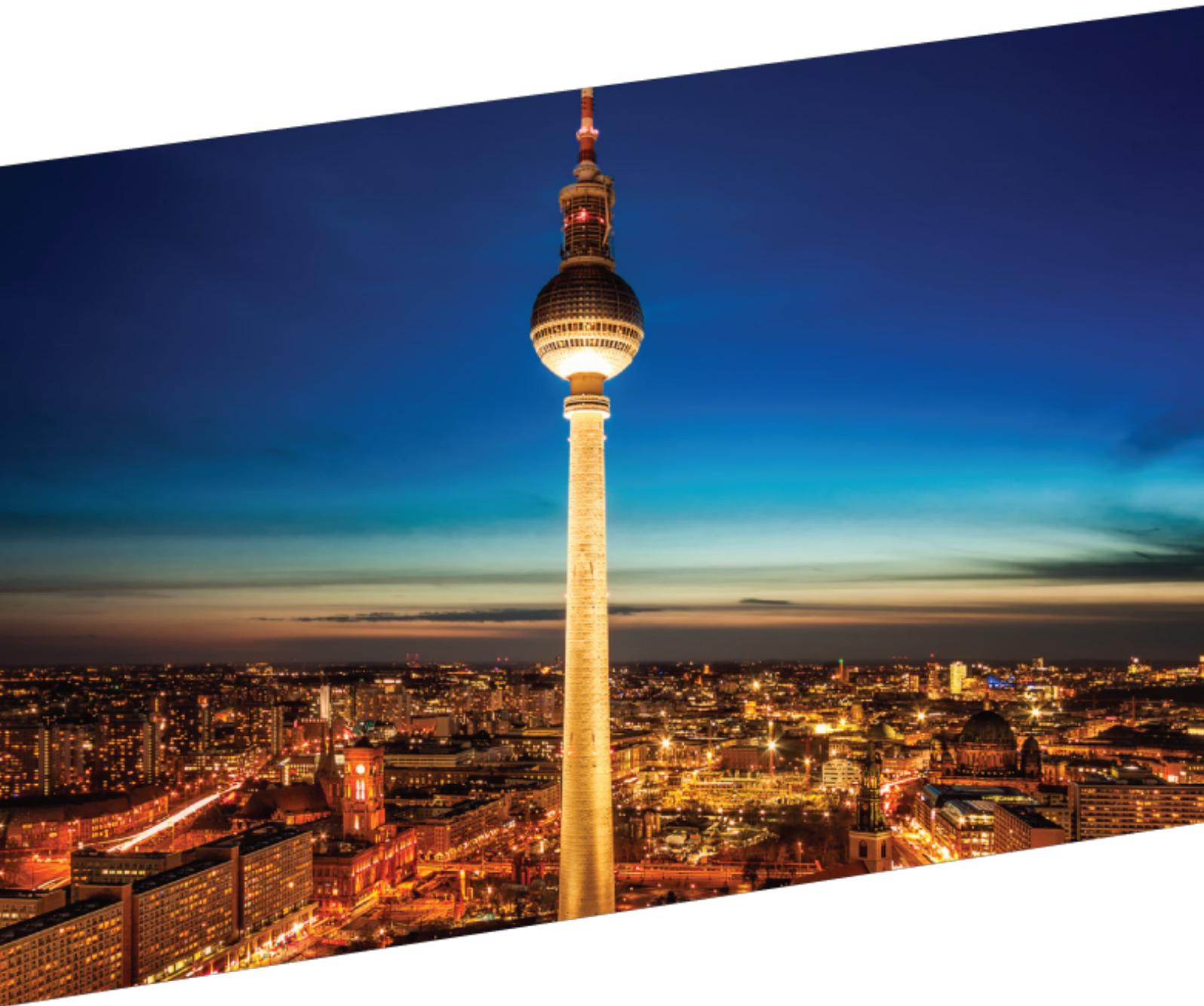

August 15-16, 2019

# Berlin, Germany

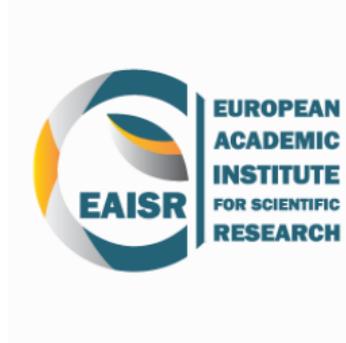

# EAISR International Academic Conferences Proceedings

## Berlin International Academic Conference  2019

### August 15-16, 2019

### Berlin, Germany

### Hosted by:

## European Academic Institute for Scientific Research

# ISSN: 2512-2940

# Table of Contents

# A Scientific Argument for Tourism Research

*Professor Kazuyoshi Takeuchi, Jissen Women's Junior College, Japan*

**Abstract:** When a discipline confronts a paradigm crisis, total disagreement and constant debate over fundamentals can be observed. And there are as many theories as there are researchers in the field. Researchers often become indulged in their own approaches and pay less attention to linkage with other research results. Coincidentally, these phenomena in the paradigm crisis resemble the present chaotic state of the study of tourism. This paper introduces the present state of the study of tourism from a different perspective and indicates an alternative research direction through extracting unscientific attitudes and conducting several thought experiments. It presents a new stance which removes widely accepted disciplinary fields from the field of tourism so as to clear the field for a genuine discussion on the raison d'être of the study of tourism.

**Keywords:** fundamental questions, enumeration, fine writing, imperfect induction, academic divide

## 1. Introduction

According to Kuhn (1996), modern physics explains that light is composed of tiny particles called photons, which are quantum-mechanical entities that exhibit some characteristics of waves and some of particles. This theory was developed by Planck, Einstein, and other scientists in early 21st century. The characterization of light is, however, scarcely half a century old. Before this theory, it was believed that light was a transverse wave motion, which is rooted in a paradigm that derived ultimately from optical writings in the early 19th century. But the paradigm with this wave theory was provided by Newton's optics, which taught that light was material corpuscles. These transformations of the paradigms of physical optics are scientific revolutions, and the successful transition from one paradigm to another via revolution is the usual developmental pattern of mature science. From what happened in the field of optics, Kuhn introduces how a field is changed when a paradigm crisis happens.

> But with continuing resistance, more and more of the attacks upon it will have involved some minor or not so minor articulation of the paradigm, no two of them quite alike, each partially successful, but none sufficiently so to be accepted as paradigm by the group. Through this proliferation of divergent articulations (more and more frequently they will come to be described as <u>ad hoc</u> adjustments), the rules of normal science become increasingly blurred. Though there still is a paradigm, few practitioners prove to be entirely agreed about what it is. Even formerly standard solutions of solved problems are called in question. [1996: pp.82-83]

In the latter stage of a paradigm crisis, as ad hoc theories come to be applied to explain the growing anomaly, the rules of the field become blurred and the solutions of solved problems are called in question, all of which resemble the present state of the study of tourism. In the review of Kuhn's discussion, Chalmers (1999) points out the characteristics of immature "pre-science".

> *It is the lack of disagreement over fundamentals that distinguishes mature, normal science from the relatively disorganised activity of immature <u>pre-science</u>. According to Kuhn, the latter is characterised by total disagreement and constant debate over fundamentals, so much so that it is impossible to get down to detailed, esoteric work. There will be almost as many theories as there are workers in the field and each theoretician will be obliged to start afresh and justify his or her own particular approach. [1999: pp.110-111]*

As a matter of course, the above-mentioned characteristics seem almost the same as the ones of the study of tourism. Firstly, there are total disagreement and constant debate over the fundamentals of the field of tourism. Secondly, there are almost as many theories as there are researchers in the field of tourism. Thirdly, researchers are obliged to start afresh and justify their own approach. Although the study of tourism has not experienced a paradigm yet or never will, its present state can be identified with the characteristics of pre-science. Considering the resemblance, this paper introduces the present state of the study of tourism from a different angle and indicates an alternative research direction through extracting unscientific attitudes and conducting several thought experiments. It presents a new stance which removes widely accepted disciplinary fields from the field of tourism so as to clear the field for a genuine discussion on the raison d'être of the study of tourism.

## 2. Fundamental Questions

When a discipline develops toward a holistic description of its research target, the quality of discipline totally depends on how its target is described with theories, as if the quality of a dinner course depends on how the ingredients are cooked with professional skills. If the quality of discipline can be represented by a linear function of its research target as a working hypothesis, the relationship between a discipline and its target can be expressed in a functional equation as shown below, regarding a discipline as $y$ and its target as $x$.

$$y = f(x)$$

When a discipline $y$ is physics, for example, the research target $x$ indicates the physical phenomenon and $f(x)$ represents the description of the physical phenomenon. And when $y$ is biology, $x$ indicates the life phenomenon and $f(x)$ represents the description of the life phenomenon. Applied with the same idea, when $y$ is the study of tourism, $x$ indicates the tourism phenomenon and $f(x)$ means a description of the tourism phenomenon.

Here is the first fundamental question. What is tourism phenomenon? There seem to be as many answers as there are researchers. The tourism phenomenon is generated by the tourist. Without the tourist, there will be no tourism phenomenon. Consequently, the tourism phenomenon consists of the subject (the tourist) and its complementary event (tourism business), which indicates that the phenomenon is an integration of "tourist phenomenon" (tourist-based) and "non-tourist phenomenon" (non-tourist-based). Borrowing Gun's model of tourism system (2002), it can be explained that the integration is comprised of a demand side and a supply side. The supply side includes the elements of attraction, promotion, information, service, and transportation. Each element involves various tourism organizations.

With a well-known windmill-shape chart, modified from Jafari's work in 1977, Goeldner and Ritchie (2012) explain that the study of tourism comprises 21 fields which are:

*Sociology, Economics, Psychology, Anthropology, Political Science, Geography, Entrepreneurship, Environmental Studies, Architecture, Agriculture, Parks and Recreation, Urban and Regional Development, Marketing, History, Law, Kinesiology, Business, Gaming, Transportation, Hotel and Restaurant Administration, Education. [2012: p.18]*

This leads to the second question. Can any or all of the 21 fields navigate researchers into the right direction to describe a holistic description of the mechanism of the tourism phenomenon? If so, or if it is on the way but takes more time, why do researchers not regard the study of tourism as a discipline or future discipline? If not, is there any good reason to require those 21 fields in the study of tourism? And if the 21 fields are not required, can it be logically recognized that they may simply confuse the field of tourism?

According to the website of Global Development Research Center, a page of Journals and Magazines in Sustainable Tourism, Urban Environmental Management introduces 127 tourism-related journals, including the fields of leisure, hospitality, recreation, destination management, heritage, environment, etc. as indicated below:

*Acta Turistica / Anatolia Journal, Annals of Leisure Research / Annals of Tourism Research / Ara (Caribbean) Journal of Tourism Research / ASEAN Journal of Tourism and Hospitality Research / Asia Pacific Analyst / Asia Pacific Journal of Tourism Research / Australian Leisure Management / Current Issues in Tourism / e-Review of Tourism Research / Enlightening Tourism. A Pathmaking Journal / European Journal of Tourism Research / Event Management / Festival Management and Event Tourism / Hospitality and Society / Human Dimensions of Wildlife / Information Technology and Tourism / International Journal of Contemporary Hospitality Management / International Journal of Culture, Tourism and Hospitality Research / International*

*Journal of Event and Festival Management / International Journal of Event Management Research / International Journal of Hospitality and Tourism Administration / International Journal of Hospitality and Tourism Systems / International Journal of Hospitality Knowledge Management / International Journal of Leisure and Tourism Marketing /International Journal of Safety and Security in Tourism - Hospitality / International Journal of Tourism and Hospitality Systems / International Journal of Tourism Perspectives / International Journal of Tourism Policy / International Journal of Tourism Research / International Journal of Tourism Sciences / International Journal for Tourism and Travel Management - Link? / International Travel Law Journal / Journal of Applied Recreation Research / Journal of China Tourism Research / Journal of Convention and Event Tourism / Journal of Convention and Exhibition Management / Journal of Ecotourism / Journal of Heritage Tourism / Journal of Hospitality Applications and Research / Journal of Hospitality Financial Management/ Journal of Hospitality and Tourism / Journal of Hospitality and Tourism Education / Journal of Hospitality and Tourism Management/ Journal of Hospitality and Tourism Research / Journal of Hospitality and Tourism Research / Journal of Hospitality and Tourism Technology / Journal of Hospitality, Leisure, Sport and Tourism Education / Journal of Hospitality Marketing and Management / Journal of Human Resources in Hospitality and Tourism / Journal of International Volunteer Tourism and Social Development / Journal of Interpretation Research / Journal of Leisurability / Journal of Leisure Research /Journal of Park and Recreation Administration / Journal of Policy Research in Tourism, Leisure and Events / Journal of Quality Assurance in Hospitality and Tourism / Journal of Retail and Leisure Property / Journal of Sport and Tourism / Journal of Sustainable Tourism / Journal of Teaching in Travel and Tourism / Journal of the Canadian Association for Leisure Studies / Journal of Travel and Tourism Marketing / Journal of Travel and Tourism Research / Journal of Travel and Tourism Marketing/ Journal of Travel Research / Journal of Tourism / Journal of Tourism and Cultural Change / Journal of Tourism and Cultural Heritage / Journal of Tourism Challenges and Trends / Journal of Tourism Consumption and Practice / Journal of Tourism History / Journal of Tourism and Peace Research / Journal of Tourism Studies - non-active - archives only / Journal of Unconventional Parks, Tourism & Recreation Research / Journal of Vacation Marketing / Journeys: The International Journal of Travel and Travel Writing / Leisure Sciences / Leisure Studies / Locum Destination Review / Loisir et Sociét / Society and Leisure / London Journal of Tourism, Sport and Creative Industries / Managing Leisure / New Zealand Analyst / Pacific Tourism Review / PASOS - Journal of Tourism and Cultural Heritage / Problems of Tourism / Progress in Tourism and Hospitality Research / Revista Turismo & Desenvolvimento (Journal of Tourism and Development) / Scandinavian Journal of Hospitality and Tourism / Studies in Travel Writing / TEOROS (Revue de Recherche en Tourisme) / Territories and Tourism / Tourism and Hospitality E-Review / Tourisme and Territoires / Territories and Tourism / The Surrey Quarterly Review / Tourism and Hospitality Planning and Development / Tourism and Hospitality Research (The Surrey Quarterly Review) / Tourism and Hospitality Review / Tourism: An International Interdisciplinary Journal / Tourism Analysis / Tourism Analyst / Tourism, Culture and Communication / Tourism Economics / Tourism Geographies / Tourism and Hospitality Research / Tourism in Marine Environments / Tourism Management / Tourism Recreation Research / Tourism Research Journal / Tourism Review / Tourism Review International / Tourism and Travel / Tourism and Travel: An International Journal of Tourism and Travel Management / Tourism and Hospitality Review / Tourismos: an International Multidisciplinary Journal of Tourism / Tourismus Journal / Tourist Studies / Tourism Today / UNLV Journal of Hospitality, Tourism and Leisure Science / Visions in Leisure and Business / Visitor Studies / World Journal of Tourism, Leisure and Sports / World Leisure and Recreation Association Journal / World Leisure Journal / Worldwide Hospitality and Tourism Themes*

And there are supposed to be far more journals written in local languages in non-English speaking countries or regions as well.

Now the third question. "What is your field of study?" This typical question can be heard from researchers who think that the field of tourism is fictitious and need to stand on an existing field. It has been a common reaction that the tourism phenomenon is too diverse or complicated to grasp. However, no matter how diverse or complicated the research target may be, there is no reason to give up analyzing the target. On the contrary, diversity or complicatedness has always encouraged humans to challenge quest for the truth. It is ludicrous to consider that physicists or biologists have given up on their research because of the diversity or complicatedness of their research target. A critical issue for researchers in the field of tourism is that many of them lack the sense of belonging to the field of tourism, so they seem to be uninterested in pursuing a holistic description of the mechanism of the tourism phenomenon.

Then the fourth question. Has the study of tourism established or will it be able to establish a distinct disciplinary field to describe a holistic vision of the mechanism of the tourism phenomenon? Some researchers may answer "yes" while some may answer "no" or otherwise "do not know". The interesting issue here is that not only "yes" researchers but also "do not know" or even "no" researchers have been submitting their papers to one of the above-mentioned tourism-related journals. It seems as if physicists who suspect or deny the domain of physics submit their papers to journals of physics or as if biologists who suspect or deny the domain of biology publish their work in journals of biology. It is just like depositing money into a bank account while suspecting the credibility of the bank. It then can be concluded that the tourism-related journals seem to be merely playgrounds for strangers from different fields who come to play as freely as possible and keep saying that they do not belong to the field of tourism. The present chaotic state of the study of tourism can be ascribed to a mixture of independently isolated research results derived from different fields.

Finally, the fifth question. Where should all the research results be archived? There are two ways – in researchers' own fields and in the field of tourism. If the first way is selected, then there is no need to set up a field of tourism. Each result is to be stored in its pertinent field. And the tourism phenomenon is to be analyzed in the existing disciplinary fields, and there will be no need for the study of tourism any more. If the latter is selected, it means that the study of tourism is, at least temporarily or as a working hypothesis, permitted to be set up with its distinct field open to the public, not as a playground for strangers.

## 3. Unscientific Attitudes

### 3-1 Qualitative versus Quantitative

There seem to be quite a few researchers who only quote the names of the authors and years of publication, sometimes as many as possible, as if the enumeration of such names and years can prove the validity of their research results. In tourism-related journals, it seems popular to employ citations with only authors' names and publication years. But solely enumerating names and years cannot support the treatise's validity. Who could examine their quotations without its passages or contexts? Even if passages or contexts are attached, it still may not prove anything unless quoted literature has come to be widely accepted among researchers. And displaying as much literature as to appeal to their knowledge in the list of references at the back of treatises is another habit which does not scientifically prove the reliabil-ity of their work. Those habits are typically observed especially when the research methods are based on qualitative analyses, rather than quantitative ones. Researchers may suffer from an obsession that the ideal research must be as rigorous as natural sciences with the assistance of mathematics. They seem sensitive about the number of quotations, especially when the method is based on a conceptual structure rather than a statistical approach, as if research without mathematic assistance is not scientific. It may lead a groundless assumption that a number of research results may impact their job-hunting for a tenured position or promotion in the organization that they belong to. Considering the reason of the enumeration is to support their own work, it is far more effective to employ broadly recognized literature or discuss the logical validity or consistency of quoted passages or contexts, rather than the simple enumeration of names and years.

There is a traditional concept called "proposition" which refers to a sentence containing a certain notion and stating a declaration. When there is a sentence "Snow is white.", the sentence is recognized as a proposition, because the word "snow" has a certain notion and the sentence gives a declarative statement. A proposition, consequently, has a characteristic of being questioned whether the statement is true or false. When "snow" indicates a limited range of meaning, such as something falling in a particular place and time, the sentence is called a "particular proposition". When it indicates a whole notion of snow observed in every place and time on earth, the sentence is called a "universal proposition". Science deals with this universal proposition. In order to reach the conclusion that the snow on earth is all white, one has to observe every case of snowing in every place from the past to the future. It is a complete enumerative inference called "perfect induction", which is technically not doable. Scientists instead apply "imperfect induction" to their research method, in which a characteristic of a whole notion (population) is inferred with a part of the notion (sample). Not only natural sciences but also social sciences, including statistical analyses, receive the benefit of this imperfect induction or incomplete enumerative inference.

Although it seems logical to think that a result from such a scientific reasoning of imperfect induction should present a truth, it does not prove any truth but only provides a probability of being a truth because the researcher has no chance to observe every case of target theme in every place and time on earth. There are many cases that a theory has once held as truth but later turned out to be a fault every time scientists discovered a new theory. The geocentric theory suggested by Claudius Ptolemy was replaced by the heliocentric theory of Nicolaus Copernicus, or Newtonian mechanics was modified by Einstein's theory of relativity, which later encountered the paradigm shift of quantum mechanics based on the uncertainty principle. The proposition that all swans are white turned out to be false when black swans were found in Australia in 1697. This is an anecdotal case of hasty generalization, which represents a probability of scientific procedures. Science takes advantages of mathematical formulae which most of the time seem accurate and more efficient than language. However, a result based on mathematical transactions always requires interpretative analyses through language to adapt for the diversity of human society. For most cases, an interpretation by language is usually applied before and after calculation.

A typical example of such cases can be observed in the theory of relativity, in which Einstein (1961) explains the validity of his theory by applying a conceptual assumption to Lorentz transformation. When an object transports from $x$ to $x'$ in a uniform linear motion with constant velocity $v$, the position of the object is given in Galilean transformation of Newtonian physics.

$$x' = x - vt \qquad\qquad x' = \frac{x - vt}{\sqrt{1 - \frac{v^2}{c^2}}}$$

*Galilean transformation*       *Lorentz transformation*    *[1961: p.37]*

When an object moves with light speed $c$, an anomaly emerges. This is when Galilean transformation has been modified into the Lorentz transformation. And when the object moves back from light speed to normal speed, the interpretation of the Lorentz transformation is required. In other words, how can Lorentz be converted to Galilean? It can be converted if the numerical value of the root becomes one. To gain one, the value of the fraction should be zero. To gain zero out of the fraction, the calculation of $v^2$ divided by $c^2$ must leave zero. Considering the fastest orbital velocity humans ever gained is calculated less than only 0.00003 % of light speed in vacuum space, the result of $v^2$ divided by $c^2$ can be interpreted as the approximation of zero, which leads Lorentz back into Galilean equation by way of combination of quantitative and qualitative processes. Researchers should be encouraged to provide quotation sources to entrust readers to the evaluation of the research results other than being indulged in the unscientific enumeration of names and years. There is no reason to be embarrassed for employing qualitative analyses.

### 3-2 Fine Writing

Bernstein (1965) states that the word "fine" in fine writing has come to mean almost the reverse of its usual meaning of superior or refined. Fine writing is ornate or overblown, hence bad writing. The above-mentioned case is recognized as "equivalence" or "elegant variation" in the field of stylistics. Equivalence refers to a kind of synonymic relation or variational device to avoid repetition of lexical items. It is associated with formal writing. Wales (2011) introduces an example of equivalence as follows:

*Film star Rocky Clint arrived at Heathrow yesterday. The macho six-footer and father of six had the fans swooning in the arrival lounge. (Underlined by the author) [2011: p.143]*

The "film star Rocky Clint", "macho six-footer" and "father of six" here are meant to be the same person. Wales explains that equivalence was found in the study of Old English poetic language in the late 19th century. The repetition of an idea expressed in a word, phrase, or clause, in different words or synonyms, is often used with grammatical parallelism. This classical custom remains in the 21st century.

Takeuchi and Nakamura (2002) analyzes the usage of three core concepts of the study of tourism such as "visitor", "tourist" and "traveler" observed in two journals – Annals of Tourism Research (ATR) and Tourism Management (TM) published between 1996 and 2000. Out of 69 issues, which is a sum of 26 from ATR and 43 from TM, they find quite a few cases of equivalence. The following two cases, which are excerpted from journals of ATR and TM.

> *In 1994 the total reached 963,995 <u>foreign and domestic tourists</u>, excluding those staying with friends and relatives. While the majority were <u>domestic travelers</u>, approximately 150,000 were <u>international visitors</u>. (Underlined by the author)*
>
> *From Annals of Tourism Research, Vol.24 (No.2)*

> *The Italian leadership in the international tourism market is confirmed by the fact that in 1935 more than three million <u>foreign travellers visited Italy</u>, whereas 1.35 million <u>visitors entered Switzerland</u>. Other competing European tourism destinations, including France, Yugoslavia and Greece, received less than 500,000 visitors. (Underlined by the author)*
>
> *From Tourism Management, Vol.17 (No.5)*

In the text from ATR, the writers express "tourists" in the first place, but it is replaced with "travelers" and "visitors" afterwards. And in the second case from TM, the writers first express "travellers" and "visited" and later replace them with "visitor" and "entered". The terms "traveler", "tourist", and "visitor" are used interchangeably as if the three of them have the same meaning. If the cases are found in literary work, it is no problem, but in academic journals, it is a problem. The English language seems to be under the spell of fine writing. It forces writers to employ aesthetic techniques of prose style. Thus, the use of equivalence in academic writing is an unscientific attitude and should be avoided for the progress of the study of tourism.

### 3-3  Academic Divide

According to Austin (2009), an estimated number of languages spoken throughout the world is 6,900, regardless of the definitive inconsistencies of local variations. It is a widely accepted idea that all the research results ever made must be shared by the people of the world. In so doing, the results must be written in a global language. English has become one of the global languages. Crystal (2003) asserts that there are basically two reasons that English has been accepted as a global language.

*The present-day world status of English is primarily the result of two factors: the expansion of British colonial power, which peaked towards the end of the nineteenth century, and the emergence of the United States as the leading economic power of the twentieth century. [2003: p.59]*

Compared to English-speaking countries, people in non-English-speaking countries seem unmotivated to employ English as a global language. This trend can particularly apply to the people of economically independent countries. In such countries, English is not crucial since the people can become well educated, obtain decent jobs, and raise children with their own local language alone. There, researchers can gain a tenured position in universities, a chance to be promoted to professorship and a satisfactory level of living without using English. The society allows researchers to work with their local languages. As a matter of course, most research results in those countries are not published in English. Knowledge is not shared with the other researchers in the world. This situation produces an "academic divide" between English-speaking and non-English-speaking countries. From the viewpoint of English speakers, research written in English can gain a chance to be the first and become a new theory, while from the viewpoint of non-English speakers, the value of research does not change no matter what language is employed. In the world of English as a global language, literature written in local languages other than English does not function as a reference, for the majority of researchers are unable to understand it. Even though the literature may be well written, it cannot be listed in references. Knowledge stagnation prevails in the academic divide. Translation may be the only hope to remove the divide, but it takes an extra cost.

The first step to analyze the tourism phenomenon was marked in the field of economics in the late 19th century, when tourist traffic was statistically heaviest at that time. There was a vital need to analyze the emerging phenomenon derived from the growing business of tourism after a new age known later as the industrial revolution began in England. Vukonić (2012) introduces the first works referred to as the beginning of expert literature written in European countries, such as Austria, Switzerland and Italy. Although those treatises triggered the dawn of the study of tourism, their contents focused on the statistic approach of tourism business. The steam locomotive was invented which enabled hundreds of people to travel safely and comfortably as a group. It changed society and this new phenomenon generated mass tourism and attracted researchers to pay attention to statistical analyses of its economic impact.

Now more than 130 years have passed ever since researchers noticed the economic impact brought by tourism business. And yet, as explained earlier, researchers have not built up a base of the study of tourism. Notwithstanding, some Asian countries provide tourism courses in universities as part of a discipline. When it comes to expressing a discipline in English, a suffix of 'ics' or 'logy' is utilized as in mathematics, physics, economics or psychology, biology, etc. The study of tourism is an exception. The term 'tourismics', 'tourismology' or even 'tourosophy' is not popular. However, it is a custom for Mandarin Chinese, Taiwanese, Japanese, and Korean to express the study of tourism as an established discipline. As Table 1 indicates, each character "学", "學", "学", or "학" means a discipline, though the idea has not been accepted. It is merely a custom of writing, which may give a false impression that the study of tourism is a discipline. Additionally, some universities in those countries are granting degrees in tourism. The different writing system may confuse society.

*Table 1: Discipline in Chinese Characters and Hangul*

| English | Mandarin | Taiwanese | Japanese | Korean |
|---|---|---|---|---|
| Discipline | 学 | 學 | 学 | 학 |
| Mathematics<br>Physics<br>Economics<br>Biology<br>Psychology<br>Study of Tourism | 数 学<br>物理学<br>经济学<br>生物学<br>心理学<br>旅游学 | 數 學<br>物理學<br>經濟學<br>生物學<br>心理學<br>旅游學 | 数 学<br>物理学<br>経済学<br>生物学<br>心理学<br>観光学 | 수 학<br>물리학<br>경제학<br>생물학<br>심리학<br>관광학 |

# 4. Essence of the Study of Tourism

## 4-1 Fence Sitters

Cohen (1979) introduces an important argument that research in the sociology of tourism should be processual, contextual, comparative and emic.

1. *Processual: Tourism is a complex process or perhaps of specific processes. There are philogenetic processes of touristic penetration, development and decline in a given area, and orthgenetic processes involving the generation and execution of individual touristic trips.*

2. *Contextual: Tourism is a process which takes place in a wide-ranging geographical, ecological, economic, social, cultural, and political context. One of the most serious drawbacks of specific studies in the field of tourism has been that this context is only rarely fully specified.*

3. *Comparative: The current writing on the sociology of tourism suffers from a lack of an explicit comparative perspective. Highly interesting analyses of specific touristic situations can rarely be used for a more general analysis because they have not been set in a comparative framework.*

4. *Emic: It emerges from the preceding discussion that it is not sufficient to study the touristic process from the outside; one has to recognize that the emic perspective not only forms, in Pi-Sunyer's (1974) term, a "separate reality", but is also of consequence for the external manifestations of touristic process.*

*[1979: pp.31-32]*

His argument indicates that the present chaotic state of the study of tourism can be ascribed to a mixture of various adjacent fields, including the aforementioned 21 fields. His first argument seems to appeal to the chaos of the field of tourism through expressing the dichotomy between two opposite genetic processes. The second reflects a typical nature of the field composed of geography, ecology, economics, sociology, culture, and politics, just as the 21 fields. The third indicates that there is no linkage among research results due to researchers' different fields. And the fourth one emphasizes a perspective of the subject (emic) other than a perspective of the observer (etic). The subject of the demand side is the tourist while the subject of the supply side is the people of destination and peripheral organizations. Regardless of Cohen's intention, his argument indicates two key words. One is a mixture of gathering fields and the other is the tourist's perspective.

Hirst (1985) claims four necessary characteristics for a discipline which can be seen in the developed forms of knowledge.

> (1) *They each involve certain central concepts that are peculiar in character to the form. For example, those of gravity, acceleration, hydrogen, and photo-synthesis characteristic of the sciences; number, integral and matrix in mathematics; God, sin and predestination in religion; ought, good and wrong in moral knowledge.*

> (2) *In a given form of knowledge these and other concepts that denote, if perhaps in a very complex way, certain aspects of experience, form a network of possible relationship in which experience can be understood. As a result the form has a distinct logical structure. For example, the terms and statements of mechanics can be meaningfully related in certain strictly limited ways only, and the same is true of historical explanation.*

> (3) *The form, by virtue of its particular terms and logic, has expressions or statements (possibly answering a distinctive type of question) that in some way or other, however indirect it may be, are testable against experience. This is the case in scientific knowledge, moral knowledge, and in the arts, though in the arts no questions are explicit and the criteria for the tests are only partially expressive in words. Each form, then, has distinctive expressions that are testable against experience in accordance with particular criteria that are peculiar to the form.*

> (4) *The forms have developed particular techniques and skills for exploring experience and testing their distinctive expressions, for instance the techniques of the sciences and those of the various literary arts. The result has been the amassing of all the symbolically expressed knowledge that we now have in the arts and the sciences.*

> *[1985: p.44]*

Based on the above-mentioned Hurst's insight, John (1997) argues the indiscipline of the study of tourism, describing the present chaotic state.

> *First, tourism studies can, in fact, parade a number of concepts. These include, for example, the destination, the tourism multiplier, yield management, tourism impacts, and tourism motivation. But these concepts are hardly particular to tourism studies. They are concepts that have started* *life elsewhere and been stretched or contextualized to give them a tourism dimension… Second, tourism concepts do not form a distinct network. They tend to be separate and atomized and indeed need to be understood generally within the logical structure of their provider discipline. They do not link together in any logical way to provide a tourism studies way of analyzing the world. Their only link is the subject of their study which is tourism… Third, tourism studies does not have expressions or statements which are testable against experience using criteria which are particular to tourism studies. Hurst gives examples of the sciences' use of empirical experimentation, and of mathematics' resource to deductive reasoning from sets of axioms. Tourism studies does not provide any truth criteria which are particular to itself but rather utilizes those criteria which are found in its contributory disciplines. [1997: pp.643-644]*

Tribe's first assertion expresses that the field of tourism consists of a mixture of various adjacent fields, just as the 21 fields or Cohen's second argument. His second one shares the same idea with Cohen's third argument. The third one implies that there are less or even no original concepts and criteria in the study of tourism and most or all of them are borrowed from different fields. Thus, from the viewpoints of Cohen and Tribe, it may be a mixture of gathering fields with no linkage among research that blocks the progress of the study of tourism, partly because there are fence sitters from different fields who come in and out of the field of tourism at their own convenience and whose research minds do not belong to the field of tourism. It provokes a suspicious thought that the present chaotic state of the field of tourism are due to a lack of integration toward a broadly accepted direction, because research results with concepts of different fields have no linkage with each other. It seems that the concepts from different fields have been impeding a genuine discussion to search for a real research target of the study of tourism. Each theory does not seem to march hand in hand to the same direction of the field while physics or biology does.

### 4-2  Thought Experiments on Tourist

If biologists are asked what the principle research target is, they may say life form. And if physicists are asked the same question, one of the answers may be substance or force. This leads to the most crucial point in this paper. If tourism researchers, who confirm that the field they belong to is tourism, not something else, are asked the same question, what can it be? The answer with the highest possibility may be tourist, because the tourist (cause) brings about the tourism phenomenon (effect), and the tourism phenomenon cannot emerge without the tourist. It is a circular argument of definition. Or if the tourism phenomenon can be divided into the demand side (tourist phenomenon) and the supply side (non-tourist phenomenon) as mentioned earlier, the demand side is to be focused, because supply does not happen without demand, which is also a circular argument.

Then comes a fundamental question. Why have humans migrated from place to place on earth ever since they climbed down the trees and started to walk on two feet? Why have people enjoyed their tours from home to destinations ever since the steam locomotive began to carry people and bring them back to the original starting point? In other words, why do people travel? Humans, just as all life forms on earth, need to maintain their life and pass their genes to the next generation. To maintain their life, they need to move from one place to another to find water and food. To move around to search for water and food, humans must perceive the environment for survival. To perceive the environment, they utilize their sensory organs to collect information from the environment through five faculties of senses – vision, audition, olfaction, gustation, and tactition. Gibson (1986) describes that:

> Let us next observe that animal locomotion is not usually aimless but is guided or controlled – by light if the animal can see, by sound if the animal can hear, and by odor if the animal can smell. Because of illumination the animal can see things; because of sound it can hear things; because of diffusion it can smell things. [1986: p.17]

Gibson's description suggests that an animal's locomotion is based on the recognition of the environment. The environment stimulates animals, including humans, to move around to collect more information of the adjacent environments. The main reason of human locomotion in the ancient stage is to obtain water and food to survive. To step into an unknown area is dangerous so humans must risk their lives when moving from their original point to the untrodden region. In the age of modern tourism, humans travel for their pleasure or other factors from their usual habitat to the destination. In other words, the tourist moves away from the ordinary to the non-ordinary, as in the definition of tourism written by Jafari (2000) in the old version of the Encyclopedia of Tourism.

> For example, tourism is defined as the study of man (the tourist) away from his usual habitat, of the touristic apparatus and networks responding to his various needs, and of the ordinary (where the tourist is coming from) and nonordinary (where the tourist goes to) worlds and their dialectic relationships. Such conceptualizations extend the frame beyond the earlier trade-oriented notions or definitions mostly devised to collect data and calculate tourist arrivals, departures, or expenditures. Significantly, it is this holistic view which accommodates a systematic study of tourism: all its parts, its interconnected structures and functions, as well as ways it is influenced by and is influencing other forms and forces related to it. [2000: p.585]

A controversial question must be asked. Who is the tourist? How should the term be defined? To discuss this matter, here are a group of thought experiments to focus on.

> Thought Experiment 1
>
> A famous tourist attraction stands next to Jane's house in her hometown (ordinary). When she visits the attraction for her assignment of college in a weekday, is she considered as a tourist? She does not leave her usual habitat.

The answer to the question depends on which definition is applied. Leiper (1993) describes that there are three definitive categories of tourists.

*Popular Notions.* This set comprises meanings used in everyday communication and con-sciousness. Two examples are reported in <u>Webster's New International dictionary</u>:

1. *One that makes a tour, one that travels from place to place for pleasure or culture, one that stays overnight usually in an inn or motel;*

2. *Tourists class: a class of accommodation (as on a passenger ship) usually less expensive and less roomy than first or second or cabin class.*

*Heuristic Definitions of Tourist.* An heuristic definition is one intended to clarify understand-ing. It may represent popular notions refined into academic concepts. Its function in this context is specifying a behavioral role (of tourist) to be studied or discussed.

*Technical Definitions of Tourist.* This category is most commonly applied in quantitative re-search Besides statistical uses, it also has legal applications, such as with immigration adminis-tration. In either case, technical definitions of tourist are usually framed from the perspective of a region or country in the role of a destination or place visited by tourists in transit. [1993: pp.540-541]

Applied by Leiper's technical definition, Jane cannot be counted as a tourist. Neither can she by Jafari's explanation with the concept of non-ordinary. As for heuristic definitions, Leiper introduces two benefits. One is that the exercise of formulating the definitions concentrates the researcher's thinking, and the other is that a definition helps readers to better understand the content. Without it, different readers are likely to infer different meanings or connotations because of the variety in popular notions. Although there are benefits in employing heuristic definitions, there is also a risk that there will be too many definitive variations created by re-searchers, as indicated earlier in this paper.

*Thought Experiment 2*

*When Jack visits Jane's house to ask her to guide around a tourist attraction next to her house (non-ordinary to him) and she does as he wishes, are both Jane and Jack considered as tourists, because the activity is the same?*

Here, this experiment focuses on the essence of the activity. Logically speaking, the essence of both behavioral patterns of Jack's and Jane's are the same, only Jack travels from ordinary to non-ordinary. If the study of tourism concentrates on locomotion, Jane should not be con-sidered as a tourist. However, it is logical to treat them as tourists since both phenomena on site are identical in nature. Both enjoy visiting the neighboring attraction.

*Thought Experiment 3*

*When a family stays in a luxurious hotel adjoining to a famous theme park (non-ordinary) and Mother just watches the regular TV program as she does every day at home (ordinary) while Fa-ther reads his favorite newspaper over breakfast as usual (ordinary), is mother's or father's rou-tine regarded as a tourist behavior?*

This experiment is a similar case to Jane's standpoint in Experiment 2. In this experiment, Jane is not concerned with locomotion from ordinary to non-ordinary but Mother and Father in this experiment are.

*Thought Experiment 4*

*When Jack on his business trip takes a break between two official meetings in a foreign country (non-ordinary) and visits a tourist attraction for five minutes in the same venue of the second meeting, is he counted as a tourist?*

The discussion point of this experiment is being at an attraction during the official hour by which his salary is calculated. Statistically speaking, he is not counted as a tourist. When he enters the country, the purpose of his visit is registered as "business", but the very moment he confronts with a tourist attraction has nothing to do with statistics. Should a conceptual discussion of the study of tourism always focus on Leiper's technical definition? Can no other arguments be permitted? The next experiment introduces another controversial case.

*Logical Experiment 5*

*When a person travels from ordinary to non-ordinary, the person is considered as a tourist. When a person stays at home (ordinary) and enjoys appreciating the vivid video images, projected on the digital screen, of a famous tourist attraction located in a foreign country (non-ordinary), is the person accepted as a tourist?*

Most researchers may not accept the person as a tourist, presumably because the concept of locomotion (physical visit) is employed as a determinant for definitive categorization. If the concept of essence (mental visit) is employed, the following case seems much easier to discuss.

*Thought Experiment 6*

*When a vision of a tourist attraction is reflected by light on the retinas of a person's eyes*

    *1)   when visiting the attraction (non-ordinary) or*

    *2)   when staying at home (ordinary),*

*is the person admitted as a tourist?*

The question in Experiment 6 problematizes the validity of the determinant to categorize a person as a tourist. Which determinant should the study of tourism take – locomotion or essence? In other words, it is the question of a physical visit or mental visit. A physical visit means locomotion from ordinary to non-ordinary while a mental visit means the essence or quality of the tourist phenomenon.

Those thought experiments are designed to draw attention to a pure discussion without concepts from different fields among tourism researchers whose research minds belong to the field of tourism, not something else. To answer the fundamental question, "Why do people travel?", the definition of tourist must be first discussed to gain a consensus among those tourism researchers. In a time when digital transactions were not yet discovered, drawing an image of an object or taking a picture of it and send the drawing or picture by mail was the only option for an initial witness to deliver its original image to other people (informants), who should visit the location of the object. Nowadays, any image can be sent or received through globally spread digital networks. The experiments indicate that there are two options for the study of tourism to deal with – a physical visit (locomotion) and a mental visit (essence). Both options share the same information of on-site phenomena while the off-site phenomena on both sides are different in terms of physical visits or mental visits.

## 5. Conclusion

It seems that most researchers consider the field of tourism as based on a variety of concepts borrowed from different fields, so such a field has less chance to establish its distinct discipline. Some researchers, however, have submitted their treatises to tourism-related journals while they say that they do not belong to the field of tourism. The field seems to consist of many alien fence sitters who come in and out of the field at their own convenience, and their research minds, they say, do not belong to the field of tourism, but different ones. Then it brings about a suspect that the present chaotic state of the field of tourism can be ascribed to a lack of unity toward a widely accepted direction, based on the research results with no linkage together, because each research employs concepts borrowed from different fields.

This paper, firstly, introduces several factors of unscientific attitudes typically observed in tourism-related journals. Then, from a viewpoint that a chaotic experience of physical optics before Newtonian age resembles the present state of the field of tourism, this paper indicates, through a group of thought experiments, several cases in which a pure discussion without loan concepts from different fields can be made. The idea of a pure discussion is simple. When the old frame of description does not well explain a newly discussed phenomenon, the description must be better modified, just as Kuhn's paradigm shift in natural sciences. If the previous fields can analyze the tourism phenomenon, the study of tourism is no longer required. It is the end of the story. If this is the case, tourism-related journals must be absorbed into their pertinent fields and researchers from different fields have no choice but "to go back home". On the other hand, if there is an argument that the field of tourism is required, it is the job for tourism researchers whose research minds belong to the field of tourism to examine the possibility whether the study of tourism can be established as a distinct discipline.

As a conclusion, it is not the diversity nor complicatedness but the fence-sitting attitude of researchers that may block the progress of the study of tourism. An attitude of depositing money into a bank account, while suspecting the bank is not a real one, is not scientific.

Just as physics or biology advances toward its destination, the study of tourism, if needed, should breathe a new life with its own concepts. If a minor field seems to be on the way to be newly born, that is a moment of truth that tourism-minded researchers gather around the tourism phenomenon and keep depositing their money into the right bank account. All the theories of existing disciplines derived from human knowledge, including physics, biology or even the study of tourism, are bound for the same direction. Every research achievement from working hypothesis to axiomatic principle marches down toward a grand theory of the universe, which can describe the mechanism of the whole world. The physical optics once experienced a difficult time, but it has finally built its own disciplinary field in the corner of physics. The study of tourism is not an exception. Who knows if the moment of truth is just around the corner?

# INVESTORS AS SIGNALS OF START-UPS' QUALITY

*Professor Shu-Jou Lin, National Taiwan Normal University, Taiwan*

Abstact: This study investigates how the characteristics of early contributors influence the success of an equity crowdfunding campaign. Specifically, it hypothesizes that institutional, proximate, or experienced contributors impose greater influence on crowdfunding success than other contributors. It also proposes that the effect of early contribution on success of an equity crowdfunding campaign is particularly significant for campaigns characterized with high technological sophistication.

# 'Chunked Lectures': A new model for conducting online lectures within Information Technology higher education

*Dr. Nick Patterson, Deakin University, Australia*
*Dr. Rolando Trujillo Rasua, Deakin University, Australia*
*Dr. Michael Hobbs, Deakin University, Australia*
*Dr. Guy Wood-Bradley, Deakin University, Australia*
*Dr. Judy Currey, Deakin University, Australia*
*Dr. Elicia Lanham, Deakin University, Australia*

| | |
|---|---|
| Aim/Purpose | The primary aim of this study in this paper aside from presenting student preference survey data and analytics data with relation to learning videos, is to address the attention span issues with our cohort of Information Technology students specifically and propose a new model (chunked lectures) for delivering online lectures within the Information Technology discipline which may manage attention span more effectively and thus may also increase comprehension and retention of knowledge. |
| Background | The digital age of learning has brought online learning to the forefront with Massive Open Online Courses (MOOCs), and degrees on platforms like FutureLearn, edX and Coursera. Further, students can enrol in campus-based courses online, and receive much learning via videos of those classes which are provided online often after the event. Thus, raising questions of where online lectures fit in the IT educational landscape and how they should be operationalised? |
| Methodology | This exploratory descriptive study used surveys to address the study aims. Convenience sampling was used to recruit the sample in 2017. The sample of interest was domestic and internationally enrolled undergraduate IT students at a major multi-campus university in Victoria, Australia. |
| Contribution | With regards to our contribution, we have provided survey results where there were 496 students eligible to participate, 144 responded positively to the invitation (29% response rate), with 116 completing all requirements for analysis; a 23% response rate. In this survey data, we collected data on self-rated attention span, comprehension and retention of knowledge through online learning. To address the problems highlighted in the results we have proposed a model for conducting online lectures to improve the process using a 'chunked' model, which aims to have a series of set breaks during a lecture in order for students to reset their attention span and come back refreshed for more learning. |
| Findings | Analysis of a survey given to Information Technology students (n=116) showed that self-rated attention span is 6-10 minutes and they (n=107) infact favour short learning videos at a rate of 74% for comprehension and retention of knowledge. Our data suggests that 21st-century student's attention span is a factor which can adversely affect the learning of knowledge in online lectures. |

| Recommendations for Practitioners | With regards to other IT teaching practitioners, we recommend them to read our students survey results and consider how they operate online or video-based learning with their respective cohorts - keeping in mind that attention span appears to be an issue among other things. |
|---|---|
| Recommendation for Researchers | The aspect of attention span in lectures is a highly debated topic as declared by Wilson & Korn (2007) where they state that attention span does vary during lectures, however, the literature does not support the notion of the 10 to 15-minute estimate. This is something our research can help to create a more solid case for in the research. In our survey data, we have n=116 students where the majority stated their self-perceived attention span is 6-10 minutes, based on learning within our discipline of Information Technology (within our discipline we historically employ a '2+2' model in each subject, which encompasses per week a 2-hour lecture and 2-hour practical). |
| Impact on Society | The impact on society from this research really comes about from discovering information about how students who study IT in the 21st century are managing with the influx of online learning and video-based learning materials. These results we presented, and our proposed model may assist in changing the way how the more digital way of teaching and learning take place. |
| Future Research | Future studies will focus on determining what students actually do when they take breaks from learning materials to ascertain what strategies facilitate a resetting of their attention span. |
| Keywords | e-learning, moocs, digital learning, higher education, information technology, teaching |

## CONCISE ABSTRACT

Lectures have been around for many hundreds of years within academia and still in use widely today. In recent years, the use of lectures has come under question in higher education with the influx of online learning and Massive Open Online Courses. This is causing universities to reconsider how lectures are handled in online learning spaces. Analysis of a survey given to Information Technology students (n=116) showed that self-rated attention span is 6-10 minutes and they (n=107) in-fact favour short learning videos at a rate of 74% for comprehension and retention of knowledge. Our data suggests that 21st-century student's attention span is a factor which can adversely affect the learning of knowledge in online lectures. Based on our analysis of our survey data we are proposing a new 'chunked' model for delivering online lectures. This model aims to address the issue of attention span and thus boosting the online learning capacity of students. In the modern era of online learning within universities, instructors should consider the changing nature of students and how they can transfer knowledge most effectively taking into account the potential issues of attention span.

CONTENT

*BODY OF PAPER*

**Introduction or background**

Traditionally, within academia, lectures involve one authority figure on a particular topic transmitting knowledge for an extended amount of time to a group of students. Whilst lectures have been considered sound educationally for generations, they have come into question in the 21st century for being ineffective and inefficient (Clark, 2014).

As with many disciplines, teaching in Information Technology (IT) brings many challenges and unique situations, especially in the digital information age. Teaching IT involves student learning of complicated technical theory on how a piece of technology works internally, as well as how to operate a piece of technology such as an application correctly. Subjects can cover a range of topics like how to program application code, how to design an interface, cyber security fundamentals and computer networking. Overarching the content, the digital age of learning has brought online learning to the forefront with Massive Open Online Courses (MOOCs), and degrees on platforms like *FutureLearn, edX and Coursera*. Further, students can enrol in campus-based courses online, and receive much learning via videos of those classes which are provided online often after the event. Thus, raising questions of where online lectures (*in this paper we define online lectures as either live video streaming of the lecture or providing video recordings of lectures after the fact*) fit in the IT educational landscape and how they should be operationalised?

The primary aim of this study in this paper aside from presenting student preference survey data and analytics data with relation to learning videos, is to address the attention span issues with our cohort of Information Technology students specifically and propose a new model (chunked lectures) for delivering online lectures within the Information Technology discipline which may manage attention span more effectively and thus may also increase comprehension and retention of knowledge. There are a number of research papers on the topic of student attention span, but most are related to observation methods as opposed to ours being self-rated attention span from the student's perspective. We are not making the claim that both are equal or comparable but just providing a different view from the student's perspective. Additionally, none of this related research is focused on Information Technology students specifically, which will be part of our contribution in terms of the self-rated attention span of this cohort.

The other contribution we will be proposing is a solution to the problem of attention span within online learning spaces, being a chunked lecturing model. Whilst the chunked lecture model we propose in this paper is primarily for Information Technology cohorts of students who are studying online, we also feel this could be utilised for other disciplines to produce more effective learning outcomes through managing attention span effectively.

The topics related to student attention span, knowledge delivery and altering lecture formats have been explored in the following related work (Bunce, Flens, & Neiles, 2010; O'Bannon, Lubke, Beard, & Britt, 2011; Schmidt, Wagener, Smeets, Keemink, & van der Molen, 2015; Varao-Sousa & Kingstone, 2015; Wilson & Korn, 2007) and will be analysed in the following literature review section.

# Literature Review

## The changing nature of lectures

A considerable amount of research has been conducted exploring alternatives to the traditional face-to-face lectures, particularly using modern online (internet capable) technologies, as reviewed by many authors (French & Kennedy, 2017; Kebritchi, Lipschuetz, & Santiague, 2017; R. & S., 2018; Venkatesh, Croteau, & Rabah, 2014). The benefits of online lectures, those classes that are streamed in real-time over the internet, have also received a lot of attention from researchers with results ranging from positive indications when combined with a flipped-classroom model (O'Flaherty & Phillips, 2015) to areas of concerns with poorly applied pedagogy (Dobinson & Bogachenko, 2018).

## Attention span

The aspect of attention span is quite important in learning and highly correlated with characteristics which can predict or influence student behaviour (Paul, Baker, & Cochran, 2012). There are many academic sources (where most is quite dated) which state that the average attention span during a lecture is roughly 15-minutes (Wankat, 2002). Several institutions have actually reduced their lectures to 15-minutes in length, based on a "consensus" that student attention span is 10-15 minutes (Bradbury, 2016). Research has shown that the number of students who actually are paying attention beings to drop massively after 15-minutes, which results in a loss in retention of lecture material (Hartley & Davies, 1978). Additional related research in memory recall after classes has shown that students remembered 70% of the information presented in the first 10-minutes of a lecture and 20% remembered information presented in the last 10-minutes (Hartley & Davies, 1978).

The aspect of attention span in lectures is a highly debated topic as declared by Wilson & Korn (2007) where they state that attention span does vary during lectures, however, the literature does not support the notion of the 10 to 15 minute estimate. This is something our research can help to create a more solid case for in the research. In our survey data, we have n=116 students where the majority stated their self-perceived attention span is 6-10 minutes, based on learning within our discipline of Information Technology (within our discipline we historically employ a '2+2' model in each subject, which encompasses per week a 2-hour lecture and 2-hour practical).

## Chunking and time-boxing

An approach to consider in relation to the delivery of lecture content is through 'chunking' or 'time-boxing' of content and activities. Whilst chunking of classes is not a relatively new concept, proposing this for University lectures and within the Information Technology discipline appears to be a novel proposition.

The three techniques we present for use within with delivery of lectures are: the Pomodoro technique (25-minute focused session + 5-minute interval) (Gobbo & Vaccari, 2008), the 'spaced learning' approach (3, 15-20 minute sessions + 10-minute interval) (Douglas Fields, 2005) and an approach drawn from behavioural and laboratory studies of Long-Term Potentiation (LTP) and long-term memory (LTM) encoding (timed pattern of 3 separate 10-minute sessions) (Kelley & Whatson, 2013).

Time-boxing is a common time management tool used to allocate blocks or boxes of time to tasks and activities. This allows the participant to focus on the allocation of work time to the tasks rather than how long they will need to get the task completed. The Pomodoro technique is one technique that has merged from the time-boxing approach. In a study of extreme programming (Gobbo & Vaccari, 2008) the Pomodoro technique was applied to build a sustained pace of learning whilst providing adequate rest breaks. The Pomodoro technique builds on the foundation of using 25-minute "on-task sessions" followed by a 5-minute "rest session" (Gobbo & Vaccari, 2008), however, at the end of the forth Pomodoro session, a longer break of around 15-minutes should be applied to allow for a greater rest period. The provision of breaks within this technique allows for the participants to focus on the tasks during the "on" sessions and be rewarded for their efforts with a mandated break.

The spaced learning approach Fields (2005) presents the importance of the breaks in the approach stating that the stimuli make use of segments or intervals to achieve long-term potentiation (LTP). Spaced learning can involve internals of minutes or days depending on the learning approach. Within a research study conducted by Fields (2005), the intervals which were applied were 10-minutes spacing's, whereas in a study conducted by Lotfolahi & Salehi (2017) the internals were on 3 consecutive days. The study conducted by Lotfolahi and Salehi (2017) found that students performed better, in regards to retention of knowledge, when they spent 5-minutes each day for 3 days on the task rather than spending 15-minutes on one singular day.

Both the Pomodoro and space learning approach use techniques which are comprised of targeted work time followed by a period of rest.

The next section will focus on the methods involved to conduct our study, where we will explain the survey tool we used and analytics revolving around video analysis.

**Methodology**

This exploratory descriptive study used surveys to address the study aims where institutional ethics approval was gained. Convenience sampling was used to recruit the sample in 2017. The sample of interest was domestic and internationally enrolled undergraduate IT students at a major multi-campus university in Victoria, Australia. Students were eligible for inclusion if they were enrolled in *Critical Thinking and Problem Solving* within the Bachelor of Information Technology and Bachelor of Cyber Security in 2017 either online or onsite, could read and understand English, provided consent through response to the survey, and were over the age of 18. There were no exclusion criteria. Students were invited to participate via email, all responses on the survey were provided anonymously and prior to summative assessments.

An online survey (through the tool *SurveyMonkey*) was developed by the researchers. Minor changes were made for clarity of information. The tool comprised of four domains: 1) student perspectives of learning via video media in the form of short pre-recorded audio-visual recordings and longer front of classroom audio-visual recordings; 2) a self-evaluation of their attention span (Likert rating scale anchored from 1 strongly agree -5 strongly disagree) during these recorded classes; 3) student preference for video duration; and 4) self-rating the importance of video recordings for completing assessment tasks, and for their comprehension and retention of knowledge. All video resources were available to students after campus-based classes via the learning management system (LMS). As a note we have not presented all the results we received from the survey, just those particularly applicable to this study.

### Video Recordings as Learning Resources

Students were given a variety of learning materials to access during the whole unit. Videos provided for, and accessed by, students in this study were consistent for the standard for delivery of the *Critical Thinking and Technology* unit in their course. Eleven face-to-face onsite classes (lectures) were audio-visually recorded and uploaded to the LMS using *Echo360*. These videos displayed PowerPoint slides plus the lecturer's narrative and were approximately 80-minutes in duration each. Students also had access to an additional 11 short premium quality learning videos recorded in a green room of approximately 2-minutes duration each, as well as 8 what we call 'front of room lecture recordings' (where a video camera was used to film the entire lecture theatre experience that included the projector with the slides and the academic instructor) where they averaged approximately 79-minutes.

### Student Attentiveness

Attentiveness was measured subjectively using self-reported items in the survey (Atwater, Borup, Baker, & West, 2017; Liebert, Mazer, Bereknyei Merrell, Lin, & Lau, 2016; Page, Meehan-Andrews, Weerakkody, Hughes, & Rathner, 2017; Roach, 2014) and objectively using the duration each student played video recordings of lectures (*Echo360* and front of room) and short learning videos. The latter data was captured as viewing time, the total duration of videos and frequency of plays by accessing the reports on the relevant unit site in the LMS in conjunction with *Kaltura*, the video hosting application. This data was collected regarding the number of unique views (users) and cumulative views of all audio-visual recordings on the LMS. As noted, longer videos were recordings of campus-based classes; short videos included an introduction to problem-solving and critical thinking, and explanations of concepts such as claims, issues, credibility and arguments. These topics were chosen to ensure key concepts that assist students to acquire key subject knowledge were readily available to support students.

To increase participation and response rate, students were prompted to respond to the survey once after agreeing to the study and *as bound by and agreed to by the Ethics Committee, the researchers did not try to increase participation rate by reminder prompts to avoid any perceived or real power imbalance between students and teachers. Similarly, no demographic data were collected to ensure the anonymity of participants.*

Responses from the surveys, video access and viewing duration data, and *eVALUate* results were analysed and reported using frequencies and percentages.

The knowledge gained from this study will help ensure future deliveries of this unit, and potentially other IT units align with the current student generation's learning needs and requirements as presented and discussed in the proceeding section.

## Results and Discussion

Based upon the analysis of results given below from our survey, its highlighted problems with attention span, comprehension and retention of knowledge of students. To provide a potential solution to these results we proposed a new model for conducting online lectures in the field of Information Technology.

This proposed model focuses on online lectures of 2 hours in duration with a set of breaks or intervals with alternate activities during this time period. Whilst we will not define any set time for breaks for instructors to use, we present three approaches to how you could 'chunk' online lectures or using 'time boxing'. These are the Pomodoro technique (25-minute focused session + 5-minute interval) (Gobbo & Vaccari, 2008), the 'spaced learning' approach (3, 15-20-minute sessions + 10-minute interval) (Douglas Fields, 2005) and an approach drawn from behavioural and laboratory studies of Long-Term Potentiation (LTP) and long-term memory (LTM) encoding (timed pattern of 3 separate 10-minute sessions) (Kelley & Whatson, 2013). There is no clear evidence as to which is best out of these approaches, we promote flexibility and suggest that instructors determine the best approach for their specific cohort of students.

We also suggest that instructors (in live streaming scenarios) can direct students to use a digital feature (if available to them in their platform of choice) whereby students can 'virtually raise their hand' (for instance used in the online class tool ("Blackboard Collaborate," 2018) to gauge if the students need a break or not. If the majority would need a break, the instructor can cease the teaching at the next available point in time and create a break interval. During the breaks, we recommend students conduct a series of activities which could be any of the following: standing or stretching to increase circulation, do a short quiz, reflect on discussed theory, and review their notes or a quick online discussion with their peers.

From our data, of the 496 students eligible to participate, 144 responded positively to the invitation (29% response rate), with 116 completing all requirements for analysis; a 23% response rate. Although 23% may be considered a low response rate, it is comparable, and indeed higher, than other standard measures of student engagement and perceptions of their learning experiences (such as *eVALUate* in 2017 where we had a response rate of 17% and 2018 a rate of 22%). As demographic data were not collected, the age, gender and per cent of online, onsite, domestic or international students in the sample cannot be reported.

### Attentiveness

Results of the survey item related to student self-perception of the longest duration they were prepared to watch a video without interruption are mapped against the duration of standard lecture recordings of 90-minutes in Figure I. As shown, students reported three peak times: 2-minutes, 6-10 (which was the highest) and 11-30-minutes. Critically, student attention span drops after 30-minutes of video play. Subjectively we suggest 10-minutes is the mean time point in a normally distributed curve depicting student self-reported attention span when watching audio-visual recordings. In terms of spread, the optimal self-perceived duration lies within the range 8 to 30 minutes. These results are consistent with two previous studies on attention span (Bradbury, 2016; Wankat, 2002).

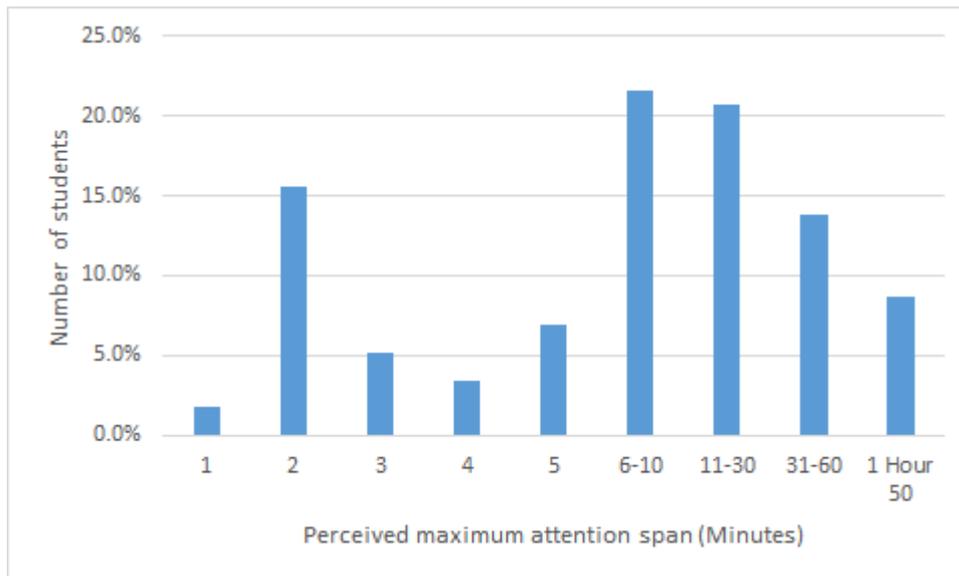

**Figure. I: Self-perceived attention span of Information Technology students with relation to video recordings (percentage 0 to 25%) n=116**

The first peak for 2-minute length video recordings may well indicate responses from 15.2% of the students who favour short premium quality learning videos of 2-minutes duration over recorded lectures of 90-minutes duration. However, no conclusive inferences can be made with regard to this postulation, given we did not study equivalently produced 2-minute and 90-minute recordings. It will be recalled that the 2–minute duration videos were high quality specially produced with high production values specifically targeted topics for assessments, rather than information sharing about topics central to the unit objectives presented in a classroom setting.

Results of the objective data taken from the LMS and *Kaltura* to indicate what actual students' actual attentiveness was are displayed in Table I. For recordings of the front of room lecture recordings (n=8), students hit the play button on average 63 times. On average, the duration of each video recording was 79.75-minutes. While 29% of those play button hits led to a video recording that was played from beginning to end, the average view time was only 22.3-minutes. Thus, actual duration was consistent with student self-perception displayed in Figure I. Regarding the short video (n=11) statistics retrieved showed on average 260.9 total plays across and these videos which lasted an average duration of 1.3-minutes. Then lastly the *Echo360* recordings (n=11) saw an average of 45.7 plays across these videos with each video total length on average being 80.2-minutes.

**Table I: Video analytics data**

| Front of room video recordings (n=8) | |
|---|---|
| Average Total Plays | 63.6 |
| Average completion | 29% |
| Average view time | 22.3 minutes |
| Average total length | 79.75 minutes |
| | |

| Short video (n=11) watchability statistics | |
|---|---|
| Average Total Plays | 260.9 |
| Average completion | 78% |
| Average view time | 1.3 minutes |
| Average total length | 2 minutes |
| | |
| **Narrated lecture slides '*Echo360*' (n=11) watchability statistics** | |
| Average Total Plays | 45.7 |
| Average completion | 57.9% |
| Average view time | Not measured by the *Echo360* |
| Average total length | 80.2 minutes |

There exist several factors affecting the view time of a video that we could not account for. For example, a student may have been interrupted during the study session or a video might have been running in the background without actually receiving the attention of the student. In both cases, the view time recorded in the LMS has little relation with attention span. Nonetheless, it is curious that the results of subjective and objective data displayed in we Figure I and Table I (front of room video recordings) are remarkably similar. That is, student attention span in practice mirrored what they had indicated as their preferred attention span in the survey.

### Implications

Considering the results of this study and published literature about student attention span when learning via video recordings of class-based onsite lectures, we suggest the adoption of teaching strategies that take into consideration this data. One approach would be for class-based teachers to adopt specific times for learning and rest breaks before recommencing teaching. For instance, we recommend using the Pomodoro technique (Gobbo & Vaccari, 2008) which provides for 25-minutes of learning and then rest for 5-minutes to maximise learning effectiveness and optimise student attention span. In this learning model, breaking the lecture delivery into a series of 'chunks', may allow the student to start off with a fresher mind after each interval and retain more knowledge, be more engaged and also boost overall grades. Additionally, instructions for watching the video via the LMS after class should include advice to limit their learning to 10-minute periods with a break that includes self-reflection and note taking about what is being learnt to consolidate their learning followed by a rest break of a few minutes to revive their attentiveness to the next section of material.

### Strengths and Limitations

The strength of this study was the design which enabled analysis of what students indicated their preferred learning attention span and the way that actually played out in the same timeframe. The relatively low response rate is an acknowledged limitation. As we could not (due to human research ethics agreement), and did not, track individual students who agreed to participate in the study by return email, and their completion of all survey data, it was not possible to specifically target those who agreed but did not complete all survey items. That said, our response rate was similar to those reported in measures of student engagement as mentioned in the results section. This study was descriptive in design, so was not powered to provide statistical significance. Further, it is a single site study conducted in one student cohort in one year, so no generalisations can be made. That said, our results may well be informative for subsequent studies in this field.

## Conclusion

Online lectures either as recordings of onsite classes or specifically developed short videos that are made available on learning management systems are a prominent feature of 21st century learning. As our data suggest students indicate they do not have or exercise an attention span to actually watch a 90-minute video of a lecture in one sitting. Thus, chunking one's learning and allowing for breaks is advised. As the optimal self-perceived duration lies within the range 8 to 30 minutes. Students should be advised to pause, reflect, take notes and rest prior to continuing. To maximise learning, the way that educators provide videoed learning material online that has been delivered previously to onsite students will need to adapt and evolve in response to the attention span of students.

We have proposed a model for conducting online lectures to improve the process using a 'chunked' model, which aims to have a series of set breaks during a lecture in order for students to reset their attention span and come back refreshed for more learning. Future studies will focus on determining what students actually do when they take breaks from learning materials to ascertain what strategies facilitate a resetting of their attention span. The chunking model presented in this paper is aimed at addressing the relatively short attention span of students that is increasingly reported in the literature for student learning.

# Inventory management practices impact on gross profit margin: A study on beverage, food and tobacco sector listed companies of Sri Lanka

*Dr. V. Sritharan, University of Jaffna, Sri Lanka*

**Abstract**

Inventory is a vital part of current assets and huge funds are committed to inventories as to ensure smooth flow of production and to meet consumer demand. Effective and efficient inventory management goes a long way in survival of a business firm. And Inventory management plays an essential role in balancing the benefits and cost associated with holding inventory. So this study investigate inventory management practices effects on gross profit margin of the companies in beverage food and tobacco sector, Sri Lanka Colombo stock exchange. A panel data from 2012 to 2016 was gathered for the analysis from the annual reports of 20 beverage food and tobacco sector firms considered. The multiple regression model was applied in the data analysis to find out the relationship between inventory management practices and gross profit margin. The variables used include inventory conversion period, operating cycle, current ratio, cash conversion cycle and gross profit margin. The results provide that inventory conversion periods, operating cycle have a significant positive relationship with gross profit margin whereas cash conversion cycle has negative relationship with gross profit margin. In addition to that, inventory management has significant impact on the profitability measures of gross profit margin.

**Keywords:** Inventory Management; Beverage Food and Tobacco sector; Gross profit margin

## 1. Introduction

Maintaining a suitable level of inventory is a key issue to firms' operational performance and inventory plays a significant role in the survival of all organizations via supporting to growth. Managing assets of all kinds can be viewed as an inventory problem; the principles used in inventories can also be applied to cash and fixed assets too. Usually, the literature of inventory focuses on production and procurement as the principal determinants of corporate inventory policy and management. In this sense, the trade-off between ordering costs and holding costs characterizes the transactions approach to inventory management represented by the Classic policy model and Economic Order Quantity (EOQ) models of inventory developed many decades ago. In recent years, as the field of operations management has developed, many new concepts have been added to the list of relevant inventory control topics (Koumanakos, 2008).

The just-in-time (JIT), material requirements planning systems (MRP) and enterprise resource planning (ERP) methods while another emerging stream of studies postulates that the characteristics of a firm's demand and marketing environments also play an important role in determining optimal corporate inventories and these are more management-oriented. Notwithstanding the theoretical or practical shortcomings inherent in these concepts and techniques, their application in real business life should have an effect in firms' performance (Koh et al., 2007).

Building on this intuition, our purpose of this study is to explore the impact (if any) of inventory management on Sri Lankan Beverage Food and Tobacco company's gross profit margin. Inventory conversion period (ICP), current ratio (CR) operating cycle (OC) and cash conversion cycle (CCC) serve as our proxy for the implementation of inventory management whereas gross profit margin was used as the profitability of beverage food and tobacco Companies of Sri Lanka.

Most literature texts declare that cost minimization or profit maximization is the main criteria for optimal inventory management, for example, an inventory manager's goal is modeled at minimizing cost or maximizing profit while satisfying customers' demands. The com-bined impact of demand planning, inventory optimization, and profit maximization can result in huge savings through reduced inventory in the system, lower clearance costs and better financial efficiencies. However, it is a large effort and it impacts a large number of users in an enterprise.

Too much inventory consumes physical space, creates a financial burden, and increases the possibility of damage, spoilage and loss. Furthermore, excessive inventory frequently compensates for inefficient and sloppy management, haphazard scheduling, poor forecasting, and inadequate attention to the process and procedures. Conversely, too little inventory often disrupts business operations, and increases the probability of poor customer service. In many cases good customers may become furious and take their business somewhere else if the desired product or service is not immediately available.

In the operations management literature, the question of how much inventory a firm should keep has been extensively studied even though there is dichotomy in the views given that inventory is both an asset and a liability. In the empirical evidence of the inventory management-performance relationship also produced mixed results. Specifically, Milgrom and Roberts (1988) and Dudley and Lasserre (1989) indicated that timely and informative customer demand data can result in improved profitability through reduced inventories. Deloof (2003) documents a significant negative relation between gross operating income and the number of inventories days for a sample of non-financial Belgian firms during the period 1992-1996, suggesting that managers can create value for their shareholders by reducing the number of inventories days to a reasonable minimum. Huson and Nanda (1995) proved that the improvement of inventory turnover (following JIT adoption) by a sample of 55 firms led to an increase in earnings per share. Additional evidence from Belgium is provided by Boute et al. (2004), who found no overall decrease of inventory ratios despite any increased focus on inventory reduction and Boute et al. (2006), who concluded that companies with very high inventory ratios have more possibilities to be bad financial performers. This is consistent with the findings of Shin and Soenen (1998), which reported a strong negative relation between the cash conversion cycle and corporate profitability for a large sample of public American firms.

Hassan, et al. (2014) examined the effect of working capital management on the performance of listed non-financial firms in Pakistan. Ordinary Least Square technique was employed to analyse data collected from non-financial firms listed on the Karachi Stock Exchange for the period 2007 to 2010. Among the independent variables used as proxy for working capital management, average age of inventory had a positive insignificant relationship with gross profit margin and return on assets.

Chen et al. (2005) in their views of inventories policies, reported that firms with abnormally high inventories have abnormally poor stock returns, firms with abnormally low inventories have ordinary stock returns while firms with slightly lower than average inventories perform best over time. Furthermore, in other study of Shah and Shin (2007) examined the empirical associations among three constructs – inventory, IT investments and financial performance – using longitudinal data that span four decades, where they conclude that reducing inventories has a significant and direct relationship with financial performance.

Contrary to the findings of the aforementioned studies, Balakrishnan et al. (1996), with the use of a small sample size though (46 firms), reported that the accounting performance of JIT adopters declines slightly compared to a matched sample of non-adopters. Further, Rotemberg and Saloner (1989) reported that a commonly identified positive association between corporate inventories and sales is greater for more concentrated industries.

Given that the results from the above few empirical studies of inventories impact on profitability are somewhat contradictory, so our study will try to shed more light to tests this issue with recent sample of Sri Lankan Beverage Food and Tobacco Companies.

The use of Sri Lankan evidence may lead to an assessment of the general applicability of inferences drawn from relevant research in different countries. To sort out the independent effects of inventories management in firms' gross profit margin (as a measure of profitability) we utilized a linear regression model estimated by this representative sector and for the financial years from 2011/2012 to 2015/2016.

## 2. Problem Statement

Many of the organizations fail to scrutinize their investment in inventory is found in the quest to maximize return on investment (Sitienei and Memba, 2016). This is unfortunate because improving the way an organization controls and manages inventory may have the greatest potential for improving the organization's bottom line (Schreibfeder, 2004). According to Temeng et al (2010), if the inventory properly managed, organizations can make potential savings but most of the organizations have continuously ignored it. For this purpose inventory should be treated as a necessary asset and requiring management.

Beverage Food and Tobacco is an essential sector to Sri Lanka and plays an important role in terms of its substantial contribution towards the growth in Gross Domestic Product (GDP) of the country (LKR 258,862 million in 2014 and annual growth rate for 2018first quarter is +06.4%), which is necessary for the country's socioeconomic growth and development. In the year of 2014, Beverage Food and Tobacco sector also was as the most performing sector in market. Recently, there has been a growing demand of Beverage food and Tobacco from hotels and tourism development activities. The government's industrial policy is to encourage investment in Food and Beverages industries as Sri Lanka has a comparative advantage. The Board of Investment (BOI) offers various incentives for investors. Also, Research Institutions conduct various programs to develop R and D facilities and Government related institutions offer training and upgrade skills of the technical staff.

As such, the increased demand has increased sales for Beverage, Food and Tobacco companies but it poses a great challenge with regards to inventory management of these companies in the country (Labour Market and Socio-economic Information Directorate (LMSID), Service Canada, Ontario). The rapid demand for Beverage, Food and Tobacco has augmented the inventory problem hence the need for effective and efficient inventory management. It is on this argument that this study aims to analyze the impact of inventory management on gross profit margin (as profitability) of Beverage, Food and Tobacco sector companies listed in the Colombo Stock Exchange (CSE), Sri Lanka.

## 3. Objectives of the Study

The objective of the study is to identify the impact of inventory management practices on gross profit margin as profitability of listed companies in beverage food and tobacco sector in Colombo Stock Exchange in Sri Lanka.

## 4. Research Questions

RQ1: Do inventory management practices have an impact on gross profit margin of listed companies in the beverage food and tobacco sector in Sri Lanka?

RQ2: What is the relationship between inventory management and gross profit margin of listed companies in the beverage food and tobacco sector in Sri Lanka?

## 5. Significance of the Study

The study is to help investors beverage food and tobacco sector firms comprehend the impact of inventory on the gross profit margin of beverage food and tobacco companies in Sri Lanka, and we hope that it helps technical individuals to employ effective control techniques in order to improve on their sector firm's works for maximized the profit. Further, this study is aimed to add knowledge to the existing literature about inventory management and gross profit margin of beverage food and tobacco sector companies in Sri Lanka.

## 6. Review of literature

Inventory is the stock purchased with the purpose of resale in order to gain a profit. It represents the largest cost to a manufacturing firm. For a manufacturing firm, inventory consists of between 20% and 30% of the total investment (Pedro Juan García−Teruel, Pedro Martínez−Solano, 2007).

Hopp and Spearman (2000) classify inventory into raw materials, work in progress, finished goods and spare parts. Raw materials are the stocks that have been purchased and will be used in the process of manufacture while work in progress represents partially finished goods.

Sekeroglu,and Altan (2014) say relationship between inventory management and profitability was analyzed with correlation and regression analysis. Accordingly, it is determined that there is a positive relationship between inventory management and profitability in eatables industry. According to the results, for the firms operating in the eatables industry, the more their inventories converted into money, the more profitability ratios included in analysis. In other words, if the firms operating in this sector sustain their inventory management policies effectively, they increase their profits.

In Ghana, Prempeh (2015) studied the impact of efficient inventory management on the profitability of manufacturing firms by using raw material inventory management and profit as variables. Cross sectional data from the company annual reports of four manufacturing firms listed in Ghana Stock Exchange were analyzed using multiple regression techniques and Ordinary Least Squares (OLS). The study found a significantly positive strong relationship between raw material inventory management and profitability.

In 2012 Gupta and Gupta concluded in their one of the study as "the efficient management and effective control of inventories help in achieving better operational results and reducing investment in working capital. It has a significant influence on the profitability of a concern thus inventory management should be a part of the overall strategic business plan in every organization."

Salawati, Tinggi, andKadri, (2012) observed the impact of inventory management on performance. They empirically studied the relationship between inventory management and firm performance on a sample of financial data for 82 construction firms in Malaysia for a period 2006-2010. They employed regression and correlation technique to analyze their findings. Their finding was that inventory management is positively correlated with firm performance. Their study focused only on general performance of the firms using financial change as a performance indicator.

Adeleke, A and Aminu (2012) found as efficient and effective management of inventories also ensures business survival and maximization of profit which is the cardinal aim of every firm. More so, an efficient management of working capital through proper and timely inventory management ensures a balance between profitability and liquidity trade-offs.

In Greece, Koumanakos (2008) carryout a study to find the effect of inventory management on firm performance in manufacturing firms operating in three industrial sectors. For this purpose, period of 2000 – 2002 considered and food textiles and chemicals sectors were used in this study. The study hypothesis of that lean inventory management leads to an improvement in financial performance of the firm was examined. The findings explored that the higher the level of inventories preserved (departing from lean operations) by a firm, the lower the rate of return. In Malaysia.

Considering all the literatures relating to this research it can see there are positive and negative relationships between profitability and inventory management. And most of the studies reviewed concentrated on conventional firm level variables such as inventory levels, demand and lead time. Sometimes the conclusions depend on the variables and population or sample which they used. However it is important to make sure that is there any relationship among these factors or not. In Sri Lankan context there are no significant researches relating to this topic. As well as most of those researches used different samples.

Furthermore according to the past studies, inventory convention period, current ratio, operating cycle and cash conversion cycle are used to measure effectiveness of inventory management for this study. In this study the Inventory management ratios are used to measure of Inventory management of the firms and therefore it is the independent variables.

## 7. Methodology

### 7.1 Data

This study mainly uses its source of data as financial statements from the selected companies' annual reports. Mainly the data from balance sheets and income statements were taken over 5 years from 2012 to 2016. All this collected financial data are in terms of Sri Lankan currencies of Rupees. The purpose of getting 5 years period of data as balanced panel data set, and the study only considered the firms that are listed in Colombo Stock Exchange (CSE).

### 7.2 Variables

According to the objective of this study, inventory conversion period, current ratio, operating cycle and cash conversion cycle are used as measures of inventory management (as Independent variable) and the gross profit margin used as a dependent variable to measure the profitability. The following hypothesis are developed based on the dependents and independents variable of this study.

### 7.3 Hypothesis

The hypothesized variables of inventory management and profitability were identified according to the review of the literatures from past scholarly works and as the reference of inventory management.

$H_1$: There is a relationship between inventory management practices and gross profit margin of listed companies in beverage food and tobacco sector in Sri Lanka.

H2: Inventory management has an impact on gross profit margin of listed companies in beverage food and tobacco sector in Sri Lanka

## 8. Model of the study

The following OLS model has used to analyses the results

$$GPM_{it} = \alpha + \beta_1 (ICP)_{it} + \beta_2 (CR)_{it} + \beta_3 (OC)_{it} + \beta_4 (CCC)_{it} + e_{it}$$

Where:

$GPM_{it}$ = Return on Asset

$ICP$ = Inventory Conversion Period

CR = Current Ratio

OC = Operating cycle

CCC = Cash conversion cycle

$\alpha$ = the Y intercept

$e_{it}$ = error

### 8.1 Analysis

Analysis was carried out in two methods of descriptive statistics method and inferential statistics method. Mainly data were collected from the audited financial reports, then sorted and analyzed by using a computerized data analysis package known as Stata12. Pearson correlation and regressions were used to measure the relationships and strength between the studied variables.

## 9. Results and discussions

Table 1 explains the descriptive statistics of the dependent and independent variables used in the study. This critical descriptive statistics examination of the dependent and independent variables discloses several issues.

**Table 1: Descriptive statistics of dependent and independent variables (2012– 2016)**

| Variable | Obs | Mean | Std. Dev. | Minimum | M a x i- |
|----------|-----|------|-----------|---------|----------|
| ICP | 100 | 63.6043 | 50.59704 | 10.12273 | 295.5786 |
| CR | 100 | 2.948528 | 5.391127 | .1167155 | 45.65175 |
| OC | 100 | 171.2038 | 108.8892 | 54.82003 | 443.4706 |
| CCC | 100 | 8.72313 | 228.4558 | -1465.523 | 1536.297 |
| GPM | 100 | 13.84257 | 15.53195 | -1.849498 | 71.36331 |

**ICP**=inventory conversion   period, **CR**=current ratio, **OC**=operating cycle, **CCC**= cash conversion cycle, **GPM**= gross profit margin.
**Source**: Results obtained from the data analysis using the statistical software package of Stata12.

This critical descriptive statistics examination of the dependent and independent variables discloses several issues. The study indicated that Profitability can be denoted by gross profit margin (GPM).

The mean (average) gross profit margin (GPM) is 13.84% to the whole sample and standard deviation is 15.53. This accounting measure (GPM) is used as profitability measure, which varies from -1.84 to 71.36 with mean ratio of 13.84%, and explains that beverage food and tobacco sector firms have an average accounting performance. The difference in gross profit margin ranged from profitability of 71.36 % (maximum value) to a loss of 1.84% (minimum value). This explores a great disparity among the firms in their profitability.

Among the measures of inventory management, operating cycle has a highest average ratio of 171.2 this means that an average number of days a company takes in realizing its inventories in cash about 171 days in beverage food and tobacco sector companies. Current Ratio (CR) has a lowest average ratio of 2.95 that's means an average short-term solvency position is 2.95 times in beverage food and tobacco sector companies.

Further the other measures of inventory management is inventory conversion   period (ICP) and cash conversion Cycle (CCC), they have an average ratio of 63.60 and 8.72 respectively.

Finally, when looking through the standard deviation (SD) measures which is helpful to know the variables' level of variation from their mean value. Here in this study explores that the most volatile variable among the examined variables is cash conversion cycle (CCC) with a S.D of 228.46 followed by operating cycle (OC) with a S.D of 108.89 then inventory conversion   period (ICP) with 50.60. Whereas the least volatile (most stable) variable is current ratio (CR) with a S.D of 5.39 followed gross profit margin (GPM) with 15.53.

*9.1 Correlation Matrix*

The correlation matrix of the independent and dependent variables is presented in Table 2 below. The results reveal that the gross profit margin has a positive relationship with the inventory management measures of inventory conversion   period and operating cycle, which are 54.00% and 41.35% respectively and those are significant at 05% level.  This is not consistent with the analysis that the lower the number of days the inventory is held in a firm before its turnover, the more the assets are utilized in the firm increasing profitability, whereas a negative relationship with cash conversion cycle, which is   -23.84% significant at 05% level (this result strengthens the finding of Takon, (2013) and Lazaridis and Tryfonidis, (2006).  In case of other inventory management measure of current ratio has low degree of strength with 06.22% and that is insignificant at 05% level.

So, we can say inventory conversion   period and operating cycle have a positive influence with the profitability measures of gross profit margin and cash conversion cycle has negative association with gross profit margin of beverage food and tobacco Companies listed in Colombo stock exchange of Sri Lanka.

This finding revel that the Sri Lankan listed firms of beverage and food sector companies' gross profit margin express a positive association with inventory conversion period and operating cycle whereas negative association with cash conversion cycle.

It implies that Sri Lankan listed firms of beverage and food sector companies' (they are small compare to firms in developed countries) inventory conversion period and operating cycle have a significant positive influence and cash conversion cycle has a negative significant influence on gross profit margin and the null hypothesis **1** is rejected in-case of inventory conversion period, operating cycle and cash conversion cycle.

This result is consistent with Shin and Soenen (1998); Lyroudi and Lazaridis (2000); Abuzayed (2012); Abuzayed (2012) and Takon (2013).

**Table 2: Correlation Matrix of the Variables (2012 -2016) and VIF**

|  | ICP | CR | OC | CCC | ROA | VIF |
|---|---|---|---|---|---|---|
| ICP | 1.0000 |  |  |  |  | 2.29 |
| CR | 0.2798* | 1.0000 |  |  |  | 2.06 |
| OC | 0.6319* | -0.1438 | 1.0000 |  |  | 1.38 |
| CCC | -0.3266* | 0.1372 | -0.3451* | 1.0000 |  | 1.21 |
| GPM | 0.5400* | 0.0622 | 0.4135* | -0.2384* | 1.0000 |  |

**Note**: ICP = Inventory conversion period; CR = Current ratio; OC = Operating cycle; CCC = cash conversion cycle; GPM = Gross profit margin, * -Significant at 05% level
**Source**: Results obtained from the data analysis using the statistical software package of StataSE12.

Normally when we are doing the regression analysis, that the statistical problem of multicollinearity issue should considered among the independent variables.

As per the recommendation of Gujarati (2003) Variance inflation factor (VIF) can be used to diagnostics of multicollinearity issues among the explanatory variables. VIF measures express that how much that the estimated regression coefficients variance is inflated due to the correlations among the predictors in the model.

The VIF values among the independent variables of Sri Lankan beverage food and tobacco sector firms are examined and tabulated with Table 2 and its appear that none of the VIF value indicates above value of 3 (cutoff value is 10) this shows that multicollinearity problem does not exist among the explanatory variables used in this study.

### 9.2 Regression analysis

Regression analysis evaluates the relationship and its strength between dependent and independent variables. Regression analysis performed and given the results below in the Table 3.

**Table 3: Effect of inventory management measures on Gross profit margin**

| GPM | Coefficient | SE | *t*-statistic | Prob. |
|---|---|---|---|---|
| ICP | .1511052 | .0396955 | 3.81 | 0.000 |
| CR | -.167911 | .2898053 | -0.58 | 0.564 |
| OC | .0113468 | .0175169 | 0.65 | 0.519 |
| CCC | -.0028653 | .0063844 | -0.45 | 0.655 |
| Constant | 2.80911 | 2.779692 | 1.01 | 0.315 |

**Note:** $R^2 = 0.3051$; Adjusted $R^2 = 0.2758$; $F_{(4, 95)} = 10.43$; Prob $> F = 0.0000$
GPM = Gross profit margin; ICP = Inventory conversion period; OC = operating cycle; CR = current ratio; CCC = cash conversion cycle.
**Source:** Results obtained from the data analysis using the statistical software package of StataSE12.

As per the results on Table 3, model has a $R^2$ value of 0.3051 that indicates the explanatory power of the dependent variables to the independent variables. It means 30.51% of the variation in gross profit margin of the beverage food and tobacco sector companies is explained through the selected independent variables; inventory management which is represent the inventory conversion period, current ratio, operating cycle and cash conversion cycle; whereas 69.49% is explained by other variables outside to this model.

The adjusted $R^2$ represent when another one variable is added to the model, how far it will explain the dependent variable. Adjusted $R^2$ is 0.2758 in this model and when added to new variable in this model, it will explain 27.58% of the gross profit margin of the companies. This indicated that our model is a predictor, which indicates that there is a positive relationship between the dependent variable (GPM) and the independent variable which is used in this model.

According to these observations of dependent variable, model values concludes that, this is a statistically good fitted model. Because to this model's F value is 0.0000 which indicates that model is significant at 5% confidence level. So the null hypothesis 2 is rejected.

**GPM = 2.8 +0.1511 ICP -0.167 CR - 0.011 OC – 0.002CCC**

## 10. Conclusions

A notable dissimilarity between the inventory management influences on profitability measure of Sri Lankan firms and developed countries' firms is that, Sri Lankan firms most probably inventory conversion period have considerable positive influence on its gross profit margin. This situation exposes that, in the Sri Lankan firm's inventory conversion period is higher.

According to the findings of operating cycle, Sri Lankan firms' operating cycle has a positive influence on gross profit margin. Further this exposes that, in the Sri Lankan firm's operating cycle's effect is considerable lower and insignificant.

Whereas findings of current ratio and cash conversion cycle have a negative influence on gross profit margin. This situation also exposes that, in the Sri Lankan firm's cash conversion cycle is considerable lower, and both of them influence on gross profit margin is insignificant.

The findings of this empirical research study propose that a number of insights from western developed theories can be convenient to Sri Lanka, further that the certain firm's specific factors which are relevant for describing inventory management and firm's profitability in Western countries are also relevance to Sri Lanka.

## 11. Recommendations

Corresponding to the findings of this empirical study, the following appropriate recommendations are stated as;

Sri Lankan beverage food and tobacco sector listed firms should try to have high inventory conversion period and high operating cycle, those can be help to make more profit (GPM). Sri Lankan beverage food and tobacco sector listed firms should try to have low degree of cash conversion cycle and current ratio; those can be help to make more profit (GPM).

## 12. Contribution to knowledge

This research study contributes to the existing literature through examining the inventory management measures which influence on gross profit margin of Sri Lankan beverage food and tobacco sector listed firms from the view of that firm's inventory management mea-sures. This study helps to understand the impact of inventory management measures with Sri Lankan beverage food and tobacco sector gross profit margin and how they affect their performance. This study findings can be helpful to board of directors and finance managers of Sri Lankan beverage food and tobacco sector firms as an output of this research study can serve as a useful database and resource material in the area of inventory management and profitability related decision makings.

# Institutional voids and social entrepreneurship: How are social entrepreneurs developing unique strategies to cope with institutional voids?

*Charles Amoyea Atogenzoya, University of Ca' Foscari, Italy*

**Abstract**

A firm's performance is shaped by its ability to effectively manage the institutional context of its operations. With the institutional settings of developing countries varying greatly from those of developed countries, institutional voids are expected to impact social enterprises (SEs) differently due to their hybridity, in comparison to their western counterparts as well as their local counterparts (traditional for-profit businesses). Thus, whilst developing strategies to cope with institutional arrangements is important everywhere, it is more critical for SEs, in particular, and entrepreneurs in general, operating in developing countries due to the underdeveloped institutional regimes that those entrepreneurs face. Yet theories and findings have been dominated by observations of, and insights derived from, developed market contexts and are not wholly generalizable to emerging market contexts. Relying on an exploratory multiple-case study approach, this study seeks to strengthen and deepen our understanding of "social entrepreneurship" in and around institutional voids such as those found in developing countries. Specifically, the study explores how social entrepreneurs are developing unique strategies to cope with formal market institutional voids. The multiple case study findings suggest that SEs adopt various strategies such as partnerships/collaborations, image management, skills and capacity building initiatives, promotion/outreach, adaptive distribution and delivery setups, etc. to cope with formal market voids. The study contributes to the literature on institutional strategizing by illustrating how firms respond to formal market voids and, more broadly, to institutional theory by providing a deeper assessment of the nature of formal market voids in developing countries from the contextual perspective of social entrepreneurs.

**Keywords:** institutional theory; institutional voids; formal market institutions; social enterprise; developing countries

# Demography Management of the Manufacturing Process in the Automotive Industry: Concept of a Holistic Lifecycle Management Tool

*Sanjiv Surendra, Daimler AG, Germany*
*Karlheinz Fischer,, Daimler AG, Germany*
*Jens Bühler, Daimler AG, Germany*
*Ingrid Isenhardt, RWTH Aachen University, Germany*
*René Vossen, RWTH Aachen University, Germany*
*Daniela Janssen, RWTH Aachen University, Germany*

*Abstract* — The German manufacturing industry is increasingly facing challenges such as globalized, volatile and heterogeneous markets due to constantly changing customer expectations, shorter product lifecycles and the shift towards mass customization, which require extreme production flexibility. Despite technological improvements, production flexibility still requires a large contribution of its major enabler: the human workforce. The demographic change, a megatrend of our time, evidently have an impact on the German manufacturing industry: increase in aging workforce, early retirement and a decline of young people. Musculoskeletal disorders, chronic diseases and further long-term health problems may increase with aging, which affect simultaneously manufacturing performance and productivity. Therefore, challenges and effects of demographic change need to be considered in an early stage of the product development process. The purpose is to ensure the best workplace conditions for every single worker and maintain his work ability in the manufacturing process. To approach this aim, a concept of demography management is developed and field-tested in the automotive manufacturing. It consists of four significant elements: (1) the production oriented management tool, (2) holistic lifecycle management starting in an early stage of the product development process, (3) active involvement of key actors and (4) joint evaluation of project status.

*Index Terms* — Demographic Change, Human Factors and Ergonomics (HFE), Product Development Process (PDP), Work Ability, Business and Change Management

## I.                               INTRODUCTION

Megatrends are global, social and macro-economic forces of development with substantial impact on people's life by changing the world rapidly in various ways and defining the future world (Kravets *et al.*, 2018). They have a timeframe from 10 to 20 years and occur at the intersection of multiple trends, which are temporally short and more specific pattern of change (Reson *et al.*, 2016).

One megatrend of our time is the demographic change. It deals with changes in the population like age structure, ratio of men to women, fertility and mortality rates as well as migration.

The current issue of demographic change is about developments of the world's population that is aging rapidly. While in 2015 only 8,5 percent of the world's population was aged 65 and over, this number is forecasted to reach 12 percent by 2030 and 16,7 percent by 2050 (He *et al.*, 2016). Significantly for the demographic change are low mortality rates in industrialized countries. Simultaneously, the proportion of employed people, aged between 20 and 64, decreases steadily owing to low fertility rates in industrialized countries.

Also in Germany, significant influences due to demographic developments are expected: the number of employed people will probably shrink to at least 23 percent until 2060 (Statistisches Bundesamt, 2015) and the average age will increase from currently 43 up to 52 until 2050 (Nerdinger *et al.*, 2016).

Hence, demographic developments have tremendous impact on German manufacturing, which is one of the key industries. Three key indicators can be used to illustrate the impact of the industry (Audretsch, 2018): (1) great manufacturing share of German gross domestic product (about 23 percent in 2017), (2) high proportion of employment (roughly 7,6 million employees in 2017) and (3) highest impact on exports (nearly 68 percent in 2016) (Statistisches Bundesamt, 2018). The majority of employees in German manufacturing are located in production areas and practice blue-collar work that is characterized by a high performance of physical operations.

The consequences of demographic developments on German manufacturing probably entail skills shortage due to high retirement and low young talent rates as well a large amount of older workforce who are in jeopardy of being impaired in their work ability (Spieß and Fabisch, 2017).

To contribute to the competitiveness, there is a need for a concept to keep all employees – both young and older people – capable of work, motivate, foster and train for new tasks and responsibilities, which requires the digital transformation process.

*A.Human factors and ergonomics considerations in the manufacturing process*

Human factors and ergonomics (HFE) is a scientific discipline to focus on work systems in which humans interact with their physical, organizational and social environment. The domain includes i.e. human capabilities, teamwork, human-machine interaction, organizational and work design (Vukadinovic *et al.*, 2018). HFE highlights the joint improvement of human performance (e.g. productivity, quality, flexibility) and well-being (e.g. health and safety, satisfaction) by a better integration of the human into the work system. This is pursued by the approach to fit environmental design primarily to the human (Dul *et al.*, 2012).

There are several scientific studies, which confirm effects of HFE on human performance in the manufacturing process  (Abubakar and Wang, 2019; Boenzi *et al.*, 2015; Kenny *et al.*, 2016; Kolus *et al.*, 2018; Ilmarinen, 2001; Govindaraju *et al.*, 2001)

Despite these coherences, there still exists from both perspectives, research and practice, a lack of awareness and insight to consider sustainably HFE in the manufacturing environment.

The research assumed for a long time in the dominant understanding of HFE as a strict health and safety issue with less contribution on performance aspects (Wells *et al.*, 2013). Furthermore, human factors' contributions and the strategic potential of ergonomics at the business level have been widely not addressed and published in business and management journals (Kolus *et al.*, 2018).

In practice, HFE considerations have not been part of the primary business strategy of orga-nizations for a long time. There have been no procedures and standards prescribed how they could be integrated into regular planning and control cycles to reach business outcomes like performance, productivity or quality (Radjiyev *et al.*, 2015; Wilson, 2014; Dul and Neumann, 2009).

The lack of this integration had several reasons: Broberg (1997) suggests poor interest from the market in ergonomically friendly products and processes, scarce financial resources and insufficient ergonomic competence. Moreover, he mentions the dominant technological orientation, the resistance of engineers in accepting knowledge and systematic procedures as well as the poor, partly missing communication between design and production department. From business management perspective, Jenkins and Rickards (2001) highlight ergonomics as a mean to prevent injuries, while providing no return on investment. Two more studies, Kariuki and Löwe (2012) and Fruggiero *et al.* add (2016) that there have been poor awareness and promotion for the importance of HFE issue, no specific law to address this domain as well as available comprehensive HFE guidelines.

Finally, all these facts result in an insufficient awareness and acceptance of HFE considerations as a health and safety tool to be positioned as part of human resource management (Dul and Neumann, 2009).

However nowadays, HFE considerations raise to gain higher attention. The number of researchers increases who complain that HFE considerations are often addressed late in the design phase and the product development process (PDP) (Reinert and Gontijo, 2017; Bligård *et al.*, 2017; Sadeghi *et al.*, 2016; Village *et al.*, 2015; Scaravetti and Montagnier, 2009; Dul and Neumann, 2009). Once strategic design decisions about the manufacturing process have been made, the resources are already determined. Afterwards, process and product changes to reduce hazardous ergonomic conditions are time-consuming, laborious and end in dramatically increasing cost (Reinert and Gontijo, 2017; Hesselbach and Herrmann, 2011).

For this reason, like Berlin and Adams, 2017 underline, the greatest possibilities are to influence HFE considerations in early product development phases. In this context, HFE considerations needs to be embedded into a specific tool that is compatible with manufacturing processes.

Accordingly, the scientific interest moves gradually from technology-centered approach in which humans are effectively considered as impersonal components, but machines and software first selected, to a more human-centered one. The human-centered framework places human characters, skills and competences at the center by not only focusing on how the human interact with technology, but also issuing why and how the technology may be of service in supporting the human work.

*B. Direction for a solution*

As a consequence of aging workforce and challenges of digital transformation, a concept is needed to develop HFE considerations into manufacturing process of the automotive industry. The approach to integrate HFE considerations should be embedded into regular planning and control cycles that are process-related compatible with usual technical topics of the manufacturing.

The focus of the concept is set on raising awareness for a more human-centered approach in manufacturing process as well as creating a communication platform for all relevant key actors and organizational departments.

The paper presents a concept for this issue, named demography management, that is developed and field-tested in a new vehicle series of one of the worldwide largest automotive corporations located in Germany.

In the following, the background of the demography management, the concept idea and implementation phase are explained and first experiences as well as initial successes are illustrated.

II.                                              USE CASE INFORMATION

### A. Prior Conditions to the Use Case

The object of presenting use case is the production business unit of one of the world's largest automotive corporation (over 90.000 employees worldwide, mostly of them employed in Germany spread over seven locations). According to its own statements, the production business unit has to face the following four challenges:

First, the age distribution reveals a surplus of elderly workers. This is partially because of people's retirement at a higher age, advances in medical sciences and a lack of younger people that is caused by a low birth rate in recent decades. In addition, it is evident that the age distribution is very heterogeneous throughout different teams – some teams in manufacturing process have a lower other a higher average age. The consequences of elderly workers are serious like the risk of simultaneous retirement of a large age group, decreasing physical and psychological working abilities as well as problems in finding successors.

Second, like competitors, the company is currently undergoing a transformation towards digitalization and higher intelligence of manufacturing processes. This action takes place on the horizontal level (across all participants in the entire value chain) and vertical level (across all organizational levels). Obviously, advances of production technologies, intelligent worker assistance systems and new product innovations like electrification change in many respects workers' job profile. Consequently, there is a qualification and competency shift from routine tasks, which are likely to be taken over by machines to creative, innovative and communicative activities that require interdisciplinary thinking.

Third, insufficient integration of HFE considerations in the PDP often result in difficulties at start of production (SOP). Those difficulties could be i.e. that the demand of qualified workers cannot be met, inadequate ergonomic workplace design that cause hazardous ergonomic working conditions and insufficient ergonomically friendly components and products.

Fourth, there exists a fear of the workforce of the undergoing transformation towards digitalization and intelligent manufacturing systems. The workforce fears to be replaced by intelligent machines and assistance systems in the near future.

These four are not only challenges of our use case related corporate, but also represent trends and challenges of the production industry (Kolus *et al.*, 2018; Hecklau *et al.*, 2016; Erol *et al.*, 2016; Zhou *et al.*, 2015; Spath *et al.*, 2013).

### B. Findings of Prior Use Case

In the first step, to overcome these challenges, a demography management evaluation concept with focus on a holistic management has been developed and prototypical tested in the corporate (Müller *et al.*, 2016). The prototypical test was in 2014. The objective is, in the long term, to provide a framework to counteract the negative consequences of demographic change. Since potential outcomes of demographical change are manifold, countermeasures include different fields of action regarding staff recruitment, personnel development, ergonomic workplace design and health management. The concept is implemented by an evaluation of the current demographical situation of a division, definition of strategic goals and then followed by derivation of suitable measures for action (Müller *et al.*, 2016).

Meanwhile, the concept of the demography management evaluation has been established and applicated in 30 divisions of the production business unit spread over a period of four years and six German locations.

Figure 1 illustrates the results of the concept application. It shows the proportion of the number of derived countermeasures for the criteria with field of actions after the roll out in 30 divisions. The analysis stresses four criteria that are assessed as crucial with great need for action: ergonomic workplace design, age-appropriate workplace design, personnel development and intergenerational knowledge transfer.

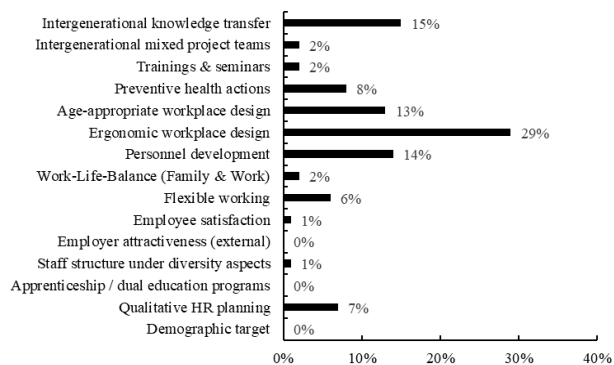

Fig. 1. Distribution (percentage) of derived countermeasures for the criteria of demography management evaluation

Out of the analysis, these criteria seem to concern a wide range of responsible persons in the production business unit, but have been neglected and insufficient considered in the past. It is to discuss, how these topics could attract higher attention within the entire organization and be dealt across the entire product lifecycle starting at an early stage of the production planning.

Therefore, as one result of the demography management evaluation, these four criteria are taken as an input to develop the concept of a demography management – the integration of HFE considerations into manufacturing process of the automotive industry – in the following chapters.

III.                                                   CONCEPTION AND IMPLEMENTATION

This chapter describes the development of a demography management to integrate HFE considerations into the manufacturing process of the automotive industry. The concept to be modelled is a further development of the demography management evaluation (Müller *et al.*, 2016) and its gained results. The demography management strives a sustainable implementation of HFE considerations at an early stage of the PDP across the entire product lifecycle management.

In the following, the criteria of the demography management, the approach of the integration in the PDP and the implementation of the concept are explained.

*A.Criteria of the Demography Management*

It is to develop a demography management model that consists of four criteria: *demographic workforce structure, intergenerational knowledge transfer, ergonomic and age-appropriate workplace design and as a further criterion production-appropriate product design.* The set of these criteria is based essentially on the analysis and experiences of the demography management evaluation (see chapter 2). In the following, the criteria are explained more in detail:

**Demographic workforce structure** aims at a greater transparency regarding the workforce's age structure starting at SOP across the entire product lifecycle. An age structure analysis serves as a systematic approach to identify in an early stage current and future personnel challenges caused by effects of demographic change (Preißing, 2014) .

For the implementation of this criterion, the following employee data should be collected and analyzed (Rosenberger, 2014): socio-demographic data (age, gender), functional data (current position and activity), structural data (operational areas), qualification data and individual data (development wishes, interests and personnel goals). Afterwards, the analysis allows conclusions regarding recruitment issues ("Are there a sufficient number of apprentice positions available?"), fluctuation issues ("What measures are taken to retain young professionals with operating experience to the company?"), arraignment of work ability ("What requirements are needed that employees can work until age of 65 or 67") and partial retirement.

**Intergenerational knowledge transfer** means the specific set of knowledge, competences and qualifications in organizations support the ability manufacture unique products. Long-lasting competitive advantages can be especially achieved by knowledge that are difficult to copy or transfer.

Generally, knowledge can be classified as explicit and tacit. Explicit knowledge is formal and structured while tacit knowledge is experiential and consisting of lessons learned (North and Kumta, 2018). The challenge is to evolve the right approach for managing knowledge in an organization.

This criterion focuses on three essential aspects: Safeguarding and transferring knowledge, identification of future needs of knowledge and exchange of specific know-how between generations.

To implement and realize intergenerational knowledge transfer, an approach consisting of five phases is applied (Kohl *et al.*, 2016): initialization, analysis, goals and solutions, realization and evaluation. The initialization phase is about data collection regarding new strategic drivers, new skills and new job profiles owing to changes in product or manufacturing process. Based on this information, in the next phase, it is to analyze if predicted skills and qualification requirements could be met for the entire product lifecycle by the existing staff. Afterwards, needs for action are identified, goals and priorities are set and solution concepts with specific measures are defined. In "realization" and "evaluation" phases, the focus is on the continuous achievement of the goals.

**Ergonomic and age-appropriate workplace design** place human beings at the center of considerations. Ergonomic workplace design is especially about human feasibility and tolerability of work activities while age-appropriate workplace design aims to maintain human work ability. Work ability means for (Ilmarinen, 2001) to put human resources on a level with workplace requirements. It is the outcome of human resources – health and functional capacities (mental, physical), qualification, attitude and motivation – related to work – work demands (mental, physical), work environment and management (Ilmarinen, 2001).

For this use-case, it is decided to apply a screening process for risk analysis (to pose health hazards) and an assessment of physical workloads across the entire PDP to aim for ergonomically friendly workplace design.

An internal, standardized two-stage IT solution method, called Ergonomic Worksystem Assessments (EAB – Ergonomische Arbeitssystem Beurteilung), is applicated for this model (Pirger *et al.*, 2017). In stage 1, first level screening is done similar to AWS light (Assembly Worksheet). Then, in stage 2, if a work process proves to be potentially ergonomic risky, a holistic evaluation of physical strains based on the Ergonomic Assessment Worksheet (EAWS) method is applied. In the two stage screening process, workload characteristics and actions like posture and body position, instances and frequency, level of action forces and environmental framework conditions are scored. A total score can be determined for every manufacturing workplace. It represents the level of risk. The level of risk is based on a three-zone evaluation system (traffic light principle) in accordance with Machinery Directive DIN EN 614-1 (Schaub *et al.*, 2013).

The three-zone evaluation system is defined as follows: Green zone (low risk, 0-25 points) workplaces are recommended and no action is needed, because the risk of illness or injury is negligible. Yellow zone (moderate risk, >25-50 points) workplaces have possible risks for illness and injury and therefore a redesign is recommended. Red zone (high risk, >50 points) points out an obvious risk of illness, which the workforce cannot be subjected to and therefore action to lower the risk is needed (Labuttis, 2015).

The purpose is to create in an early stage a greater transparency of workplaces, which are ergonomically risky to redesign or lower the risk before strategic design decisions are made (Dul and Neumann, 2009).

**Ergonomic product design** sets the focus on design of ergonomically friendly and assembly-appropriate components. The topics of interest are i.e. handling of components and parts, industrial joining, component accessibility and adjusting, which influence the workers operations (Lotter and Wiendahl, 2012).

An early transparency of ergonomically unfriendly product design issues enables redesign options. In the use-case, vehicle prototypes are assembled, ergonomic product design issues, are analyzed and assessed with EAB-It tool. Those who were assessed as critical and not ergonomically friendly, are dealt by a task force.

*B. Integration in the Product Development Process (PDP)*

In this paper, the model of demography management is to be integrated into the PDP. This approach has the following strategy: Firstly, it meets the idea of embedding HFE considerations into a specific tool that is compatible with manufacturing processes. The PDP is a standard operating procedure that is a proven process and esteemed by the entire organization. Secondly, it describes tasks and responsibilities for every key actor for the whole product lifecycle – independent from organizational department and a period of nearly 10 years.

The PDP is a stage-gate model and describes the entire workflow from a first idea for the new product to its manufacture and market launch. The PDP is divided into distinct stages, every stage is separated in specific quality gates (milestones) where process continuation needs to be decided jointly by a steering committee. Moreover, it illustrates process chains as well as roles and tasks of the participating departments (Weber, 2009).

Figure 2 illustrates common patterns of different corporations PDP representing an indus-try-wide accepted structure in automotive development.

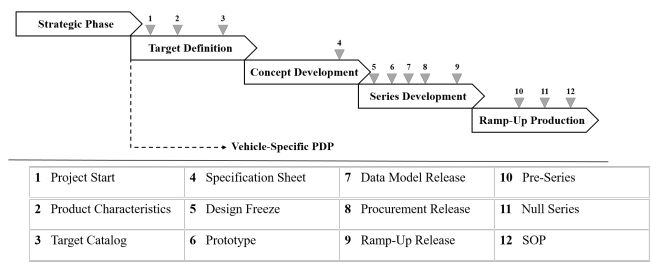

Fig. 2. Generic product development process for automotive industry (according to Göpfert *et al.*, 2017)

In the use-case, the demography management process starts at quality gate 5 (design freeze). The term "design freeze" marks the end of the design phase at which technical drawings are signed off and released to production. At this point, the first vehicle prototype is produced, field testing takes place, the plant layout is determined and concepts for materials handling and workplace designed. The design freeze is therefore the right time to initiate the HFE aspects in manufacturing process with reference to the criteria of the demography management.

In accordance with the idea of this paper, figure 3 shows the modified product development process with the integration of demography management at quality gate 5 (Design Freeze).

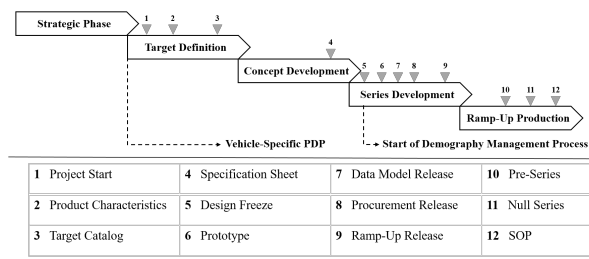

Fig. 3. Modified product development process with the integration of demography management process

### C. Concept Implementation

For implementing the demography management in the organization, participative ergonomics techniques are used in this paper (Burgess-Limerick, 2018). There are nine elements that the literature proposes as a key to a successful evidence-based approach (van Eerd *et al.*, 2010):

1. Gain support from the entire organization
2. Establish an advisory or steering committee to guide the  process
3. Hold management accountable for providing adequate resources
4. Create small teams of right people to drive the realization
5. Provide the right people with the organizational knowledge
6. Provide the right people with the right ergonomics training
7. Establish clear responsibilities
8. Deal all decisions as group consultation who are      involved in the assessment and solu-

tion process

9.Encourage active communication among all participants throughout the entire project

In the following, it is described how the process flow of the demography management is carried out by these elements. In the demography management, there is a high involvement of key actors who have first-hand experience with HFE issues like workplace design, human resource development and manufacturing process. This includes participants from various departments (e.g. production planning, assembly, human resource, R&D and works council) and hierarchical levels (e.g. managers, supervisor, workers) within the entire organization (correlate with elements 1, 4, 5, 6).

A project team steers the demography management process and reports regularly the project status to the member of the board of management for production (CEO). The CEO holds all organizational authority for decision-making and resources for his business unit and is interested in a smoothly running manufacturing process (correlate with element 2). In this way, it raises organizational and management attention that are indispensable for a successful project realization (correlate with element 3).

The project team is characterized by its neutrality and focuses on a properly functioning process, while other partners generally operate on departmental oriented goals. Besides, the project team is willing to be aware for the project content, guide them through the process as well as coach them in their tasks, authority and responsibilities. In case of conflicting interests between partners, the project team is willing to support expedient and fair solutions (correlate with elements 2, 7).

An autonomous committee, named "Round Table" serves to gather common requirements, agree on goals, give recommendations for action, derive measures and assesses jointly the implementation progress (correlate with elements 8, 9). "Round Table" is in use case a metaphor for a place where representatives of different departments gather without hierarchy to discuss divergent interests, manage conflicts and reach compromises satisfactory for all parties.

Small working groups implement the recommendations for action and take measures. Their meetings take place between "Round Tables" and personnel composition rely on the specific topic to be elaborated.

IV.                    FIRST PROJECT ACHIEVEMENTS AND RESULTS

The pilot for this use-case started at quality gate 5 of the PDP as described in the paper. After two years of pilot testing and one year before SOP, a short summing up can be made. In a review workshop, first project experiences with the demography management process, lessons learned as well as project achievements are presented, discussed and assessed together with the project partners. In the following, the mainly results of the review workshop are pointed out.

Firstly, the assembly of physical vehicle prototypes with great focus on ergonomic working activities and processes create transparency in an early stage of the PDP. First level screening as described in capture 3B is applied and components or work processes, which prove to be potentially ergonomic risky, are marked as critical and listed. By analyzing those points in detail, discussing in workshops with specialists, searching for alternative solutions in other vehicle series or finding new approaches, some successes could be achieved.

Secondly, thanks to workforce development analysis, employees with specific work limitations could be identified. Those are in general difficult to match with a suitable ergonomic workplace, especially after SOP. By the identification to that early stage in PDP, many of them could be matched to ergonomically friendly workplaces which definitely will reduce problems after SOP.

A further indicator for success is the promotion of a more human-centered corporate culture, that is appreciated by the project partners and noted by several soft facts:

The project raises awareness of key actors for HFE considerations in vehicle series projects, an understanding for demographical challenges within the corporation and risks of further missing insights of a more human-centered PDP.

Key actors of the "Round Table" pursue different goals. While works council and human resource department mainly focus on workers and working conditions, designer und planner rather have production outcomes like costs, time and quality in mind. Therefore, it is a great success that all project partners agree on one common commitment concerning project realization and target achievement by persuasion through open and controversial minded discussion and confidence-building.

The project creates greater transparency in tasks, competences and responsibilities of every partner within the product development process. The transparency leads an understanding why the key actors do certain things as they do and motivates the key actors for multidisciplinary cooperation to achieve project targets.

The "Round Table" is a committee for an open, fair and constructive dialogue between partners of different hierarchical levels and departments. The willingness of communication has spread meanwhile also to other corporate projects and initiatives.

Although HFE considerations still have today a subordinate role in the PDP, since technical issues and business outcomes are more important, there is an ongoing change of awareness. It gains more and more attention in dialogues and workshops. Besides, HFE topics are more than before in the focus and mindset of the corporation and its management. The information flow to the management plays an essential role. For those who have previously neglected HFE considerations, the attention of management exerts pressure to achieve positive ergonomic outcomes.

V.                                    DISCUSSION AND OUTLOOK

The present work sets out a methodology for mastering and improving the PDP with focus on a more human-centered approach at an early stage in the process across the entire product lifecycle. The concept of the demography management as presented in this paper is a further development of the demography management evaluation of (Müller *et al.*, 2016). It is field-tested in the vehicle development process of a model series, but can be applied on other heavy product manufacturing (such as aircraft, ship or turbine). The model illustrates both the criteria to define common targets, derive measures and assess project status as well as the methodology for project implementation.

There are several questions remaining open regarding the project progress of the use case until SOP: firstly, will it be possible to solve all the issues that are assessed as ergonomically critical with the EAB-It tool? Secondly, will an ergonomic landscape of the assembly lines be finalized until SOP – a transparency about all workplaces assessed to the three-zone evaluation system and target achievement of no red zone workplace? Thirdly, will all workers be matched to a suitable workplace with regard to required skills and qualifications? Fourthly, will the common commitment for a successful project realization be maintained at that current high level until SOP? Finally, will all the project partners do continue to see a corporate-wide and intra-departmental benefit in the concept of demography management.

The demography management presents a change management process by creating an impact on the corporate PDP, the management (corporate strategy), the corporate culture (attitude and values) and key actors (tasks and responsibilities). Lewin defines a three-step process for a successful organizational change that involves the stages of unfreezing, moving and refreezing (Lewin, 2016).

Unfreezing means disrupting the balance of driving and restraining forces by creating a sense of urgency for a change.

Moving is where the both forces, driving and restraining, are modified to shift the equilibrium to a new level. This stage presents the real transformation effort where standards are implemented and processes modified.

Refreezing, the third stage, involves reinforcing new behaviors in order to maintain new levels of performance and avoid regression. The steps "unfreezing" and "moving" are fulfilled successfully by the use case. Now it is about to implement the demography management long-term in the organization. Therefore, it is important to implement a sustainable follow-up process of the demography management in further vehicle series of the corporate.

One of the lessons learned of the project partner is that the methodology, presented in this paper, is helpful and an approach to center HFE aspects at an early stage of the PDP. However, it has to be considered, if the start of the demography management at quality gate 5 (Design Freeze) – as presented in this paper – is too late for greater process and product changes to reduce hazardous ergonomic conditions. It is also possible to think about a start of the demography management before the specification sheet is determined (quality gate 4). The advantage of this is in the longer time period to influence the product design with regard to ergonomically friendly aspects.

The demography management centers the human with the aim to maintain his work ability for the future. But the paper ignores a greater focus on the needs and desires of the workforce and how far these could be change in the future.

In addition, the model sets the focus on the megatrend of demographic change and the ongoing digital transformation process. Besides demography and digitalization, there are still further megatrends with impact on the manufacturing process in the near future. Therefore, it should be considered in the next step, in how far the demography management as presented in this paper can be applied or need to be developed to meet further challenges and megatrends.

# Relationship between Art and Society: An Analysis

*Dr. Asma Kazmi, Aligarh Muslim University, India*

### A. Art: Meaning and Significant Features:

It is not easy to define the term 'Art.' However, the term 'art' derives from the Latin *"ars"* (stem art) which although literally defined means "skill, practical knowledge of a subject". The word art is often used in a more specialized way to mean fine arts. Such as painting, pictures, writing novels or composing music and so on.

There are various views regarding the word "art" and there is no single definition of art. It has been defined in different ways. It means 'branches of learning'. It is an activity of the human mind indulging for its own sake under certain aesthetic principles. Art is a medium through which an artist can express his emotion, feelings and thought, which he wishes to share with other members of his society. In a broad sense, we can say that it is a creative product of man. In reality, every thinker or aesthetician interpreted art according to his own understanding. Among the primary concept employed with regard to art are play, illusion, irritation, beauty, emotional expression, imagination, intuition, wish fulfilment, pleasure technique, sensuous surface, form, function and the like. Some of these terms refer primarily to the creation of art, other as to the art object, and still others to the act of application.[1] [1].

Generally, it is said that Art is the creative activity that reflects reality in artistic images embodying man's aesthetic attitude to it. Painting, sculpture, fiction, theatre, cinematography, drama, music, choreography, poetry and the like are some of the important forms of art. It is also a form of social consciousness as well.[2] [2].

Art has been defined in many ways keeping in view of the time and requirement of its utility. Generally, it is the expression of spirit or the concrete idea. It has two essential forms:

---

[1] 1.Elsen E.Elbert, Purposes of Art (II ed.), Hall, Renehart and Winston, Inc, New Yourk

[2]

1. Pure knowing; and 2. Action or practical.

The second form is again divided into two actions: 1. Practice 2. Making.

The art or Making is further divided into two types: 1. Craftsmanship 2. Fine Art

Fine Art is distinguished by the nature of its function- to make beautiful rather than useful things.[3] [3]. G.T.W. Patrick has beautifully described art, kinds of art and its importance in social groups. He says,

When intelligence and skill are expanded in productive activity, we speak of this activity as art, if this activity is of such a kind as to lead to the production of objects of utility; we call it useful art; if it leads to the production of objects of beauty, we call it a fine art. There is no difficulty whatever in explaining the presence in a social group, group of the useful arts, such as blacksmithing or shoe- making or weaving since they serve the vital needs of the people. But so far as we can see works of fine arts serve no such need. And yet, not only at the present time but as far back as we can go n the history of man, the creative work of the artist, whether in music, painting, poetry, sculpture, architecture or decoration is discovered in every social group.[4] [4].

Every artist, every poet, and every thinker has defined and interpreted art in light of his or her own personal world-view and value- system as well as conditioning and cultural upbringing. Art satisfies our aesthetic sensibilities as well as our inner creative urge. According to Parker art is the provision of satisfaction through the imagination, social significance and harmony.

**B. Society: Meaning and Significant Features:**

While defining society, it can be said that 'society is a group of people, living together and sharing their required needs, feelings and sentiments, joys and sorrows  as well exercising their normative science and so on. Generally, we can say that the totality of all kinds of joint human activities, which are historically depending on the economic and political system, may form a society. People, in general, living together in organized communities with laws and traditions.

---

In the context of Art, society has to be prepared to support the ability of an artist. Now, in the following pages, we shall discuss about the relationship between Art and Society.

## C. Relationship between Art and Society:

Art is the mirror of society, very true it is said, for the art of any time period reflects the spirit of any stage of the economic and political system. The social and historical background directly influences the Artist of every period:

> Society needs Art, for the development in culture and development of humankind. An artist through his rich bountiful and overwhelming feeling give a new way for the development of society, using other brush or chisel attempted to give these abstract ideologies embedded in the through the use of different teachings of philosophers and point give a new remarkable grace and dynamism blossomed.[5] **[5].**

In the modern conception, the definition of art is altogether changed that is simply an expression of feelings. Artist himself is a man of society that he has to convey his feeling naturally to the others. But the real function of art is to express feeling and transmit understanding.

The Greek thinkers previously recognized this notion and thus, Aristotle rightly observed that the purpose of drama was to purge out emotions as a clue of peace and tranquillity.

There is undoubtedly an unchanging interrelationship between artist and community. The artist is a member of society where he from his childhood to his maturity is brought up and experiences are surroundings. But the individual character of the artist work depends on his will to form that signifies the personality of the artist. The value of art is based on his individuality, time and circumstances. The artist's relation with his audience will be an integral part of his aesthetic experience and that is why the artist does his best to know well (truth) traditionally we believe that art imitates life.

An artist represents, through his creativity, a scene of various dimensions on a canvas. Whether he is a sculptor, dramatic, photographer or writer all these are known as mime or imitation. According to an Indian thinker, Radha Kamal Mukherjee, each distinctive society has its own characteristics art form.

---

[5]

The relationship between Art and Society varies all over the world and during different eras. According to Croce, the work of an artist is an activity of the artist's consciousness. Thus, the concept raises a problem concerning the artist's relations with his audience. Artist communicates his experiences to others and in order to do this, he must have means of communications:-

Art is generally the emotion of the artist when the emotion is given expression it becomes an art. But what generally expects in the work of art is the personal element.[6] [6].

Tolstoy defines art as to invoke in oneself a feeling one has experienced and having evoked it in oneself. By the means of movement, line, color, sound, or forms expressed in words so to transmit that feeling that other experience.[7] [7].

Thus, we can say that art and society are closely related. There are certain crucial problems in the society and the artist, being a member or agent of his society tries to pinpoint those problems with the help of various art forms and also tries to give solution those problems.

Moreover, art must be for the sake of society and an artist has a crucial role to play. With the help of his art medium, he can make his society good and also tries to mould in the right direction. Art is connected with every aspect of man- social, political, religious, and ethical or economics. A sensitive art always works for the service of humanity irrespective of any man- made consideration. Thus, art and society are interdependent.

With the help of his art medium, a sensitive artist looks deeply what is going wrong in his society and tries to give a right direction to it. Art is an emotional expression in which a man transmits his own feelings to others. Tolstoy says that art is false art if it does not communicate the inner feeling of members of a social group or society.

Man has a social consciousness and he exercises his normative (ethics) in society. Art and society closely related. The question of the relationship of art to social life has al-

---

ways played as a vital role in all literature that has reached a certain level of development. There are two ways, which are directly opposed to each other:

Some thinkers and artist have said and still say "Society is not made for the artist, but the artist for the society." They hold that art must promote the society. They hold that art must promote the development of human consciousness and the improvement of the order. Briefly, art is primarily a "social force" and that the artist has a social responsibility and that the artist has a social responsibility.

Other thinkers reject this view and say that "Art is an end itself and to turn it into a means of achieving some other ends, however noble, is to lower the dignity of a work of art". In brief, we can say that "art must be for the sake of art" and not for anything else.

Man is a social being. He has his existence in society and rightly called a product of society. Man cannot live and lead his life alone. He has social contracts to survive himself in society and satisfy his own ego. It is mainly the social milieu that makes or mars human individuals' life.

With the passage of time, like other disciplines art has also changed and is still changing its tone, spirit, and techniques, according to the need of the day it is constantly moving and fulfills the requirements of the individual and society. Art is basically a social force and 'voice' and therefore, therefore highlights, likes primitive society, the problems of contemporary society.

## Role of an artist in Contemporary Society

It depends upon artist how he visualizes different problems of his society. For the artist, art is the only medium. Art is that social fence which unable him to pinpoint, certain important issues and problems of his society.

In present scientific and technological society, the role of artist becomes important and effective to give a right direction to his society all norms are changing; art gets new method and technique with the help of new scientific outlook.

In own social structure, science and technology brought new changes. To tackle societies must have a scientific mind and a better understanding of the implication of his different art form. If he knows the system of technological system advancement in the field of art he can help his society can provide better ways for the welfare of its different aspects. Art is an effective medium to make the norm of society well and suitable.

Social forms and intellectual needs had undergone a profound change by the end of the 18th Century, and it fell to the ethnic doctrine of the 19th century to assign art its vital place in a new world. In Germany, where social and economic progress was hampered by the political calamity of the worse empire, philosophers thought of the coming era in terms of metaphysics and they remained busy with restoring the concepts for the better understanding of art. The study was used for a metaphysical explanation of the universe and god. But the people of the (then) West Germany, on the other hand, very consciously perceived the advent of the modern world in a more concrete way. They raised such problems as were concerning with human society. Various problems were put to deal with the social life of man. The English and French students of art asked: What is the legitimate place and purpose of art in social life? In France, where after the revolution, the structure of society had to be rebuilt from the bottom, the bottom, the problem was felt deeply. The artists maturely affected by all these circumstances and problems and answered the too impetuous claims of society with the instrument of art.

Thus, we can say that Art had been a source of inspiration throughout human history. It may be said that art is a medium through which man expresses his emotions, attachments, and desires. In every age and at every stage art has been playing an important role in the individual and social life of individuals. It deals with every aspect of human life. Art occupies a significant place in contemporary society also. Man being a social person has to contribute to his society and his medium is art. So his very being is known thought his society.

# EVALUATION OF WOMAN ECO-ENTREPRENEURSHIP IN CONTEXT OF MOTIVATION FACTORS AND BARRIERS IN RURAL AREAS :A CASE STUDY OF DUZCE

*Dr. Pinar Gültekin, Duzce University, Turkey*

*Dr. Yaşar Selman Gültekin, Duzce University, Turkey*

**Abstract**

Turkey Entrepreneurship Strategy and Action Plan (2015-2018) 's overall aim is "to promote a culture of entrepreneurship in Turkey, to create a strong ecosystem and develop entrepreneurship" is. Within the scope of the plan, developing and implementing a sustainable support system in priority thematic areas such as women's entrepreneurship, young entrepreneurship, eco-entrepreneurship, social entrepreneurship and global entrepreneurship and in general areas are among the priority targets.

In this study, 8 different woman eco-entrepreneurship have been interviewed in entrepreneurship examples from rural were evaluated. The motivation factors and barriers related to the service types of entrepreneurs were determined through the interviews. As a result, suggestions are made to increase women's employment and women entrepreneurship and to understand and support the importance of women in eco- entrepreneurship.

**Key Words:** Düzce, Eco-Entrepreneurship, Woman, Motivation, Barriers

# The Role of Gender in Forest Engineering Profession in Turkey: Women's Role and Importance

*Dr. Yaşar Selman Gültekin, Duzce University, Turkey*

## Abstract

As in different disciplines in Turkey, forest engineering profession in addressing the problems caused by the difference in gender and does not have a sufficient number of studies demonstrating the importance of women in the profession. It is thought that women are unable to provide the required qualification of forest engineering profession in Turkey, because it is dominated by women are not adequately acknowledged in understanding male-dominated profession and unable to develop his career.

The aim of the study is to reveal whether women who have an important role in the forest engineering profession have been discriminated against because they are women. The opportunities and constraints encountered by the female forest engineers in their working conditions in the public and private sectors have been put forward, and the methods developed for success in the profession and the obstacles to private sector women entrepreneurship have been evaluated. Scope of work; In the forest engineering profession, 21 female forest engineers were interviewed in fields such as academicians, engineers, nursery operators, etc. The information obtained has been interpreted by integrating with the legal administrative framework and professional requirements and the importance of female forest engineers in the profession has been emphasized.

**Key Words:** Forest Engineering, Gender, Woman, Equlity

# Brexit Spillover Stock Market Contagion and Volatility: Evidence from an Emerging Market

*Ashraf U Bhuiyan, East West University, Bangladesh*

Global economic and political shocks significantly affect financial markets in general and emerging financial markets in particular. In this paper, we attempt to identify and analyze the channels of transmission of stock market turbulence and the impact of international contagion on emerging markets. Using daily data from January to December 2018 on FTSE100 index as a proxy for the UK market and Nifty50 index as a proxy for the South Asian market, we try to quantify and analyze the price movements and volatility spillover effects between the two markets, around the Brexit period . We divide our study period into two parts – pre- and post-BREXIT referendum – to compare the statistical significance of the transmission shocks and spillover effects between the post and pre crisis period. The results show that there indeed exists short run and long run dynamic and co-integrating relationships between the two markets; the correlation being amplified in the shorter horizon during the BREXIT crisis, both by the uncertainty of the referendum results and information adjustment due to the post- crisis shocks. The quantile regression results also evidently put forth the important global& domestic variables influencing various levels of the post-BREXITNifty returns.This evidently indicates that the FTSE 100 index affects the Nifty index and acts as a regulating gauge for the information transmission and its dynamics across the two markets.

Keywords: Volatility spillover, Market integration, Contagion, BREXIT, Emerging markets, Financial Crisis

JEL Codes: C32, E44, F36, G10